\documentclass[aps,prx,twocolumn,superscriptaddress,floatfix]{revtex4-2}

\usepackage[table]{xcolor}

\usepackage{graphicx}
\usepackage{dcolumn}
\usepackage{bm}
\usepackage{soul}
\usepackage{amsmath}
\usepackage{siunitx}
\usepackage[english]{babel}
\usepackage{xr}
\begin{document}

\title{\textbf{Emulating 2D Materials with Magnons}}

\author{Bobby Kaman}\email{Contact Author: kaman3@illinois.edu}
\affiliation{Department of Materials Science and Engineering and Materials Research Laboratory, The Grainger College of Engineering, University of Illinois Urbana-Champaign,  Urbana, Illinois 61801, USA}
\author{Jinho Lim}
\affiliation{Department of Materials Science and Engineering and Materials Research Laboratory, The Grainger College of Engineering, University of Illinois Urbana-Champaign,  Urbana, Illinois 61801, USA}
\author{Yingkai Liu}
\affiliation{Department of Physics and Institute for Condensed Matter Theory, The Grainger College of Engineering, University of Illinois Urbana-Champaign, Urbana, Illinois 61801, USA}

\author{Axel Hoffmann}\email{Contact Author: axelh@illinois.edu}
\affiliation{Department of Materials Science and Engineering and Materials Research Laboratory, The Grainger College of Engineering, University of Illinois Urbana-Champaign,  Urbana, Illinois 61801, USA}





\date{\today}

\begin{abstract}
Spin waves (magnons) in 2D materials have received increasing interest due to their unique states and potential for tunability. However, many interesting features of these systems, including Dirac points and topological states, occur at high frequencies, where experimental probes are limited. Here, we study a crystal formed by patterning a hexagonal array of holes in a perpendicularly magnetized thin film. Through simulation, we find that the magnonic band structure imitates that of graphene, but additionally has some kagome-like character and includes a few flat bands. Surprisingly, its nature can be understood using a 9-band tight-binding Hamiltonian. This clear analogy to 2D materials enables band-gap engineering in 2D, topological magnons along 1D phase boundaries, and spectrally isolated modes at 0D point defects. Interestingly, the 1D phase boundaries allow access to the valley degree of freedom through a magnonic analog of the quantum valley-Hall insulator. These approaches can be extended to other magnonic systems, but are potentially more general due to the simplicity of the model, which resembles existing results from electron, phonon, photon, and cold atom systems. This finding brings the physics of spin waves in 2D materials to more experimentally accessible scales, augments it, and outlines a few principles for controlling magnonic states.
\end{abstract}


\maketitle


\section{Introduction}
The isolation of single layers of carbon in 2004 essentially gave birth to the experimental field of two-dimensional (2D) materials \cite{novoselov_electric_2004}, and immediately prompted further study of the quasi-relativistic nature of charge carriers in graphene \cite{katsnelson_chiral_2006, li_observation_2007, guinea_energy_2010}. The catalog of 2D materials grew to include a zoo of electronic phases, including semiconductors \cite{podzorov_high-mobility_2004} and insulators \cite{novoselov_two-dimensional_2005} at first, but later more exotic interacting phases \cite{ugeda_characterization_2016,cao_unconventional_2018,zeng_thermodynamic_2023}. Among the later species to join the zoo are 2D magnetic materials \cite{han_graphene_2014,huang_layer-dependent_2017,gong_discovery_2017}. For example, ferromagnetism resembling both Ising-like and XY spins can be found within the family of chromium halides \cite{liu_thickness-dependent_2019, bedoya-pinto_intrinsic_2021} and antiferromagnetism can be found in nickel halides \cite{ liu_vapor_2020,bikaljevic_noncollinear_2021}. However, much of this research goes beyond the equilibrium states of 2D magnets, and focuses also on novel magnetization dynamics. 

In magnetically ordered solids, including 2D materials, the low-energy spin excitations are usually wave-like collective excitations called spin waves (whose quanta are magnons -- the terms will be used interchangably). Magnons are promising as charge-free information carriers for low-power computing devices in a field of study referred to as magnonics \cite{kruglyak_magnonics_2010, serga_yig_2010, chumak_magnon_2015}. Unique applications of magnons include signal guiding using broken time-reversal symmetry and inference tasks that exploit their inherent nonlinearities \cite{wang_inverse-design_2021, papp_nanoscale_2021, zenbaa_universal_2025, korber_pattern_2023}.
Magnons in 2D and layered magnets bear similarities to their electronic counterparts and can host, for example, Dirac points \cite{schneeloch_gapless_2022}, topological gaps \cite{chen_topological_2018}, and flat bands \cite{riberolles_chiral_2024}. Although some properties of these states are attractive, their applications seem distant because of their high frequency (on the order of THz) and limited number of experimentally readily accessible probes.

One strategy of controlling magnons is given by magnonic crystals, analogous to the more well-known photonic crystals \cite{krauss_two-dimensional_1996, joannopoulos_photonic_1997, nikitov_spin_2001}. By patterning a magnetic material into an artificial lattice, one may modify dispersion relations and open band gaps at the Brillouin zone (BZ) edges \cite{chumak_current-controlled_2009, wang_nanostructured_2010, tacchi_band_2011}. This is a powerful control scheme for magnon transport that may find application in next-generation microwave and information technologies \cite{chumak_magnonic_2017}. In this study, we outline a general strategy towards mapping between a few 2D materials and magnonic crystals supported by magnetization dynamics simulations and simple models. Understanding this mapping allows engineering of magnonic band structures using principles from 2D electronics. This brings some of the physics of 2D materials to more accessible frequency and length scales in a system that allows the addition of tailored defects and other spatial inhomogeneities.  Therefore we provide a simple experimentally feasible platform that allows to probe directly the emerging physics of 2D materials, while also providing a design strategy for functional devices at microwave frequencies.

\section{Spin Dynamics Simulations}
Phenomenologically, classical magnetization dynamics are described by the Landau-Lifshitz-Gilbert [LLG] equation \cite{landau_theory_1935, gilbert_phenomenological_2004}: 
\begin{equation}
    \partial_t\mathbf{M}=-\gamma\mu_0\mathbf{M}\times\mathbf{H}_{\text{eff}} + \dfrac{\alpha}
{M_S}\left(\mathbf{M}\times\partial_t\mathbf{M}\right),
\end{equation}
where $\mathbf{M}$ is the magnetization with magnitude $M_S$, $\gamma$ is the gyromagnetic ratio, and $\mathbf{H}_{\text{eff}}$ is an \textit{effective} magnetic field having contributions from externally applied fields, the exchange interaction, crystalline anisotropies, the demagnetizing field, etc. This first term gives rise to the well-known precessional motion of the magnetization $\mathbf{M}$. $\alpha$, the contribution of Gilbert, is a phenomenological damping parameter that weakens precession over time. A magnetic material may be thought of as being composed of many volume elements, known as \textit{macrospins}, whose internal magnetization profiles are roughly uniform and who each individually follow this equation of motion. It is possible to perform simulations based on this principle, in which $\mathbf{H}_{\text{eff}}$ is calculated for each macrospin and time is evolved in steps. This approach to numerical simulations, referred to as 'micromagnetism,' \cite{brown_micromagnetics_1963} is rich in phenomena and is particularly helpful in studying magnetization switching, complex geometries and ground states, and nonlinear/soliton dynamics. Here, we use the GPU-accelerated micromagnetics program \textsc{MuMax}3 \cite{vansteenkiste_design_2014} to study spin waves. [Fig. \ref{fig:Geom}(a,b)]. We choose realistic material parameters of the prototypical magnonics material yttrium-iron garnet (YIG), which is widely used in magnonic devices due to its low dissipation. See SM sec. \ref{Supp-sec:S_Geometry} for details \cite{Supplemental}.

\subsection{Geometry}
Any periodic magnetic structure can be considered a magnonic crystal. However, geometries like arrays of disks face a disadvantage in being coupled only by the relatively weak magnetic dipole-dipole interaction. One approach is to embed magnetic components in a different magnetic matrix so the components may be loosely thought of as resonators coupled through spin waves in the matrix \cite{puszkarski_magnonic_2003, krawczyk_magnonic_2013, centala_compact_2023, yang_flatbands_2025}. Another approach uses the so-called "anti-dot" lattice, in which a continuous thin film is patterned with an array of holes \cite{neusser_anisotropic_2010, ulrichs_magnonic_2010, bali_high-symmetry_2012, gros_phase_2021, wang_observation_2023}. The anti-dot lattice is different because it cannot generally be thought of as a clear set of coupled resonators. For instance, a lattice with vanishingly small holes should approach the behavior of a pristine film. Note that similar geometries have been explored in the artificial spin ice community, mostly for their interesting ground states \cite{qi_direct_2008, gartside_realization_2018}.

\begin{figure}
    \centering
    \includegraphics[width=1\linewidth]{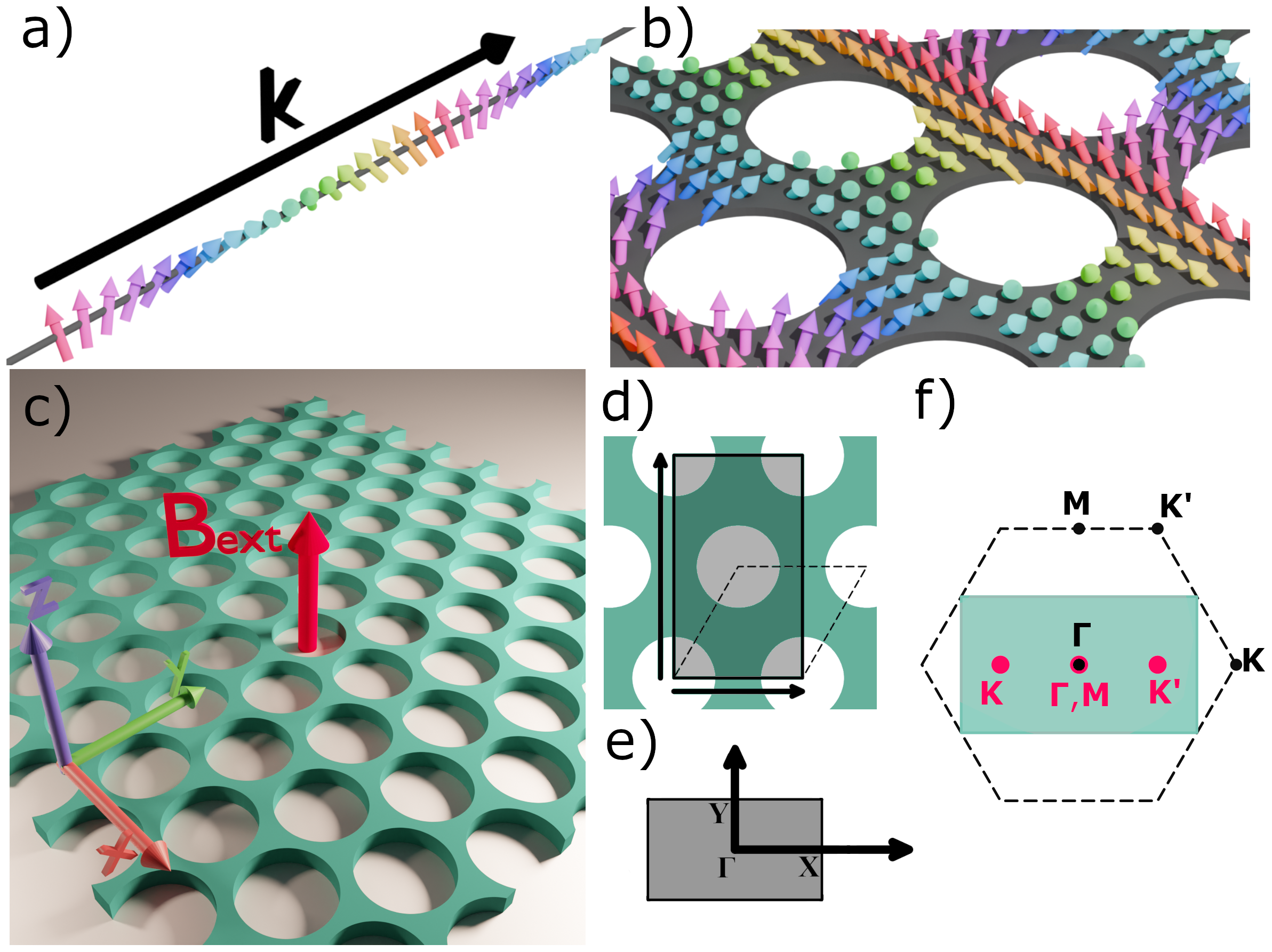}
    \caption{\label{fig:Geom}
    Spin waves in (a) 1D and (b) 2D. Each arrow indicates a macrospin and its color indicates the direction of the in-plane magnetization and hence the phase of the precession. System geometry: (c) The hexagonal anti-dot lattice with a perpendicular field, leaving a film similar to a honeycomb lattice. (d) (shaded) our rectangular supercell, with orthogonal lattice vectors $333 \text{ nm } \times 589 \text{ nm}$, and (dashed) the typical unit cell of the hexagonal lattice. (e) Corresponding Brillouin zone (BZ) for the rectangular unit cell, with a few special points $\mathbf{X}$ and $\mathbf{Y}$ marked. (f) relationship between special points in the typical BZ (dashed, black letters), and their placement in the rectangular BZ (shaded, colored letters). 
    }
\end{figure}

In general, a spin wave dispersion relation depends upon the magnetic ground state. Even for pristine thin films, the dispersion is anisotropic for an in-plane magnetization due to the dipole-dipole interaction \cite{stancil_magnetostatic_2009}. To emulate the isotropic dispersion that is natural for electrons in 2D, we choose an out-of-plane (OOP) ground state by applying an external field $B_{\text{ext}}|| \hat{ z}$ sufficient to saturate the thin film [Fig. \ref{fig:Geom}(c)].  Alternatively, one could also assume magnetic thin films that have perpendicular magnetic anisotropy. We choose anti-dot spacings of 100's of nm to balance experimental feasibility with large frequency band widths. Due to the rectangular shape of micromagnetic cells, it is tremendously convenient to work with a rectangular unit cell. We define a rectangular supercell containing two primitive cells [see Fig.~\ref{fig:Geom}(d)]. As a result, the BZ is halved and contains double the number of bands [Fig.~\ref{fig:Geom}(e)]. Note that special points in the hexagonal BZ ($\mathbf{K}$ and $\mathbf{K'}$, for instance) now fall inside the rectangular BZ, as shown in Fig.~\ref{fig:Geom}(f). These are technically new points, but we will refer to them as $\mathbf{K}$ and $\mathbf{K'}$ for simplicity.

For numerical simulations, a rectangular unit cell tens of micromagnetic cells in width is tiled tens of times; these two sizes determine real-space and inverse-space resolution, respectively. The simulation contains only one micromagnetic cell in the $\hat{z}$-direction -- this is a good approximation for very thin films, in which the exchange interaction penalizes $\hat{z}$-direction texturing and standing waves (see supplemental Sec.~\ref{Supp-sec:S_Geometry} \cite{Supplemental}). A film thickness of $15\text{ nm}$ is chosen to ensure accuracy. After the structure is relaxed to its OOP equilibrium state with periodic boundary conditions in the $x$ and $y$ directions, excitation can be applied to probe the magnonic band structure. 

\subsection{Excitation and analysis}
An excitation meant to probe a band structure should couple to many wavevectors and frequencies -- in other words, it should resemble a $\delta(\mathbf{r},t)$-function. Therefore, within a small region, a radio frequency ({\em rf}) pulse is applied of the form $B_x(t)=B_0\frac{\text{sin}\left[2\pi f(t-t_0)\right]}{2\pi f(t-t_0)}$, whose spectrum contains all frequencies between $-f$ and $+f$. $B_0$ is chosen to be small ($ \approx 0.1 \text{ mT}$ or less) to make sure we stay in the linear regime. We time-evolve the system under the LLG equation and record the magnetization state $\mathbf{m}(t,x,y,z)$ as a unit vector. Precession about $\hat{z}$ is assumed, so the complex field:
\begin{equation}
    \Psi\equiv m_x+im_y
\end{equation}
 is a convenient representation that will be used frequently. This field can be interpreted as a wavefunction in some sense. Assuming circular precession and neglecting damping, eigenmodes (by definition) should keep their shape $|\Psi(\mathbf{r},t)|^2$ and time-evolve only by a phase:  $\partial_t \Psi(\mathbf{r},t)=i\omega \Psi(\mathbf{r},t)$ for a specific angular frequency $\omega=2\pi f$. To give a relevant example, the linearized Landau-Lifshitz equation, including an exchange interaction of strength $D_{\text{ex}}$ and an externally applied field $H_z(\mathbf{r})$ can be rewritten (see appendix \ref{sec:App_Schrodinger}) in a Schr{\"o}dinger-like form assuming a $\hat{z}$-polarized ground state\cite{landau_magnetism_nodate, herring_theory_1951, lim_ferromagnetic_2021}:
\begin{equation}
 \partial_t\Psi(\mathbf{r},t)=i\gamma  \mu_0\left(H(\mathbf{r})-D_{\text{ex}}\nabla^2\right) \Psi(\mathbf{r},t)
\label{eq:Schrodinger-like}
\end{equation}
This equation is not used here \textit{except} to provide an intuitive interpretation of mode profiles as wavefunctions. In the simulations of this study, the dipole-dipole interaction is included, which, of course, deviates from this form. The $\Psi$ representation is not necesssary, but is convenient and will be enlightening in drawing parallels to electrons in real 2D materials.

The time evolution of the magnetization after excitation is written in terms of $\Psi(\mathbf{r},t)$. Because it is a result of the excitation, we will refer to it as a response. Bloch's theorem can be used to extract the magnonic band structure from the response: the signal is folded into $\Psi(t,x,y,i,j)$ where $(i,j)$ are unit cell indices and $(x,y)$ label positions within the unit cell. Then, the signal is Fourier transformed along the $t$, $i$, and $j$ axes, resulting in the complex amplitude $\Psi(f,x,y,k_x,k_y)$ associated with frequency $f$ and crystal momentum $(k_x,k_y)$. This approach is similar to Ref.~\onlinecite{feilhauer_unidirectional_2023}. 
\section{Magnonic band structures}
\begin{figure}
    \centering
    \includegraphics[width=1\linewidth]{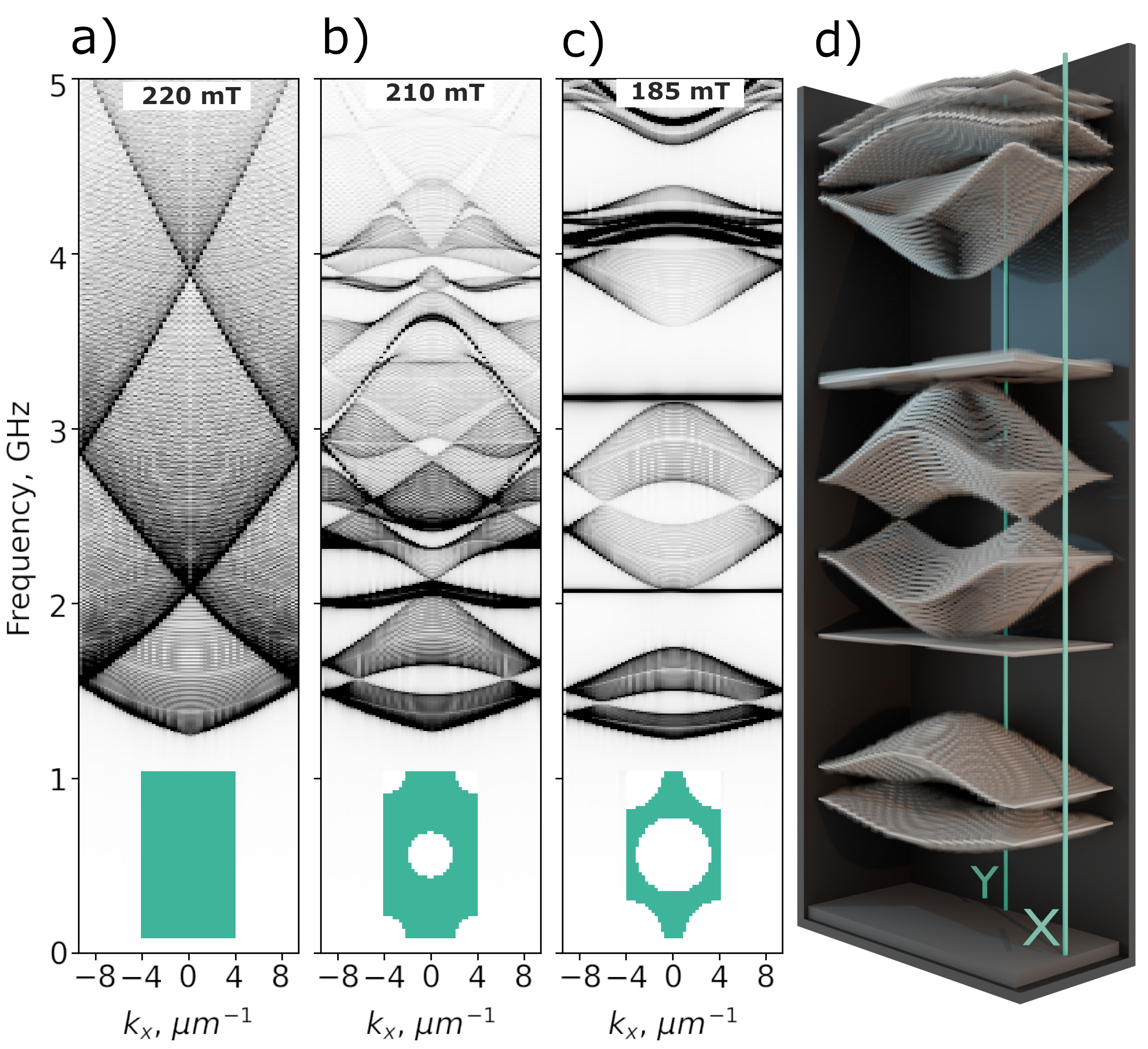}
    \caption{The effect of hole diameter $d$ on YIG films with lattice parameter $a=333\text{ nm}$, thickness $t=15\text{ nm}$. Band structures are plotted as projections in the $(k_x,f)$ plane for: (a) $d/a=0.0$ (no holes; equivalent to the free spin wave spectrum), (b) $d/a=0.4$, and (c) $d/a=0.8$. Rectangular unit cells appear as insets. $B_{\text{ext}} ||\hat{z}$ is varied to keep the lowest-frequency mode constant. (d) a 3D volumetric plot of the projection in (c), more clearly showing the isotropic character of flat bands. The $d/a=0.8$ geometry is the main subject of this study. To be more precise, the plots are the magnitude of the response $|\Psi(f,x,y,k_x,k_y)|$ to a $\delta$-like excitation, displayed as projections on to only a few axes: $(k_x,f)$ for 2D, or $(k_x,k_y,f)$ for 3D.}
    \label{fig:Bands}
\end{figure}

Though our analysis methods are general, we focus mainly on the YIG hexagonal anti-dot lattice. A few magnonic band structures are plotted in Fig.~\ref{fig:Bands}. After summing over the unit cell $(x,y)$, the magnitude of the response $|\Psi(f,k_x,k_y)|$ is a function of frequency and wavevector, so it can displayed as a projection on the $f,k_x$ plane or plotted in 3D as volumetric information. 
As expected, a continuous film [see Fig.~\ref{fig:Bands}(a)] has a magnonic band structure equivalent to the zone-folded representation of a free dispersion -- the "nearly-free-magnon." Small holes [see Fig.~\ref{fig:Bands}(b)] significantly change the dispersion but do not open spectral gaps. However, large holes open spectral gaps and bring about some curious features [see Figs.\ref{fig:Bands}(c) and (d)]: near $1.5\text{ GHz}$, the band structure resembles that of graphene. Near $2.5\text{ GHz}$, the band structure seems to have a second set of  Dirac points, and additionally has a few flat bands that meet dispersive bands at the $\mathbf{\Gamma}$ point. The nature of the transition between Fig.~\ref{fig:Bands}(b) and ~\ref{fig:Bands}(c) is not obvious; see supplemental Sec. \ref{Supp-sec:S_MoreBandStructures} \cite{Supplemental} for an expanded figure. 

Because of its curious features, this third geometry appearing in Fig.~\ref{fig:Bands}(c) is the main subject of this study. Our primary goals are to:
\begin{enumerate}
\item Point out and understand the curious features which naturally occur in this anti-dot lattice
\item Propose potential uses of its unique properties
\item Use understanding from the field of 2D materials to engineer these excitations in ways not possible with natural van der Waals systems

\end{enumerate}

\begin{figure}
    \centering
    \includegraphics[width=1\linewidth]{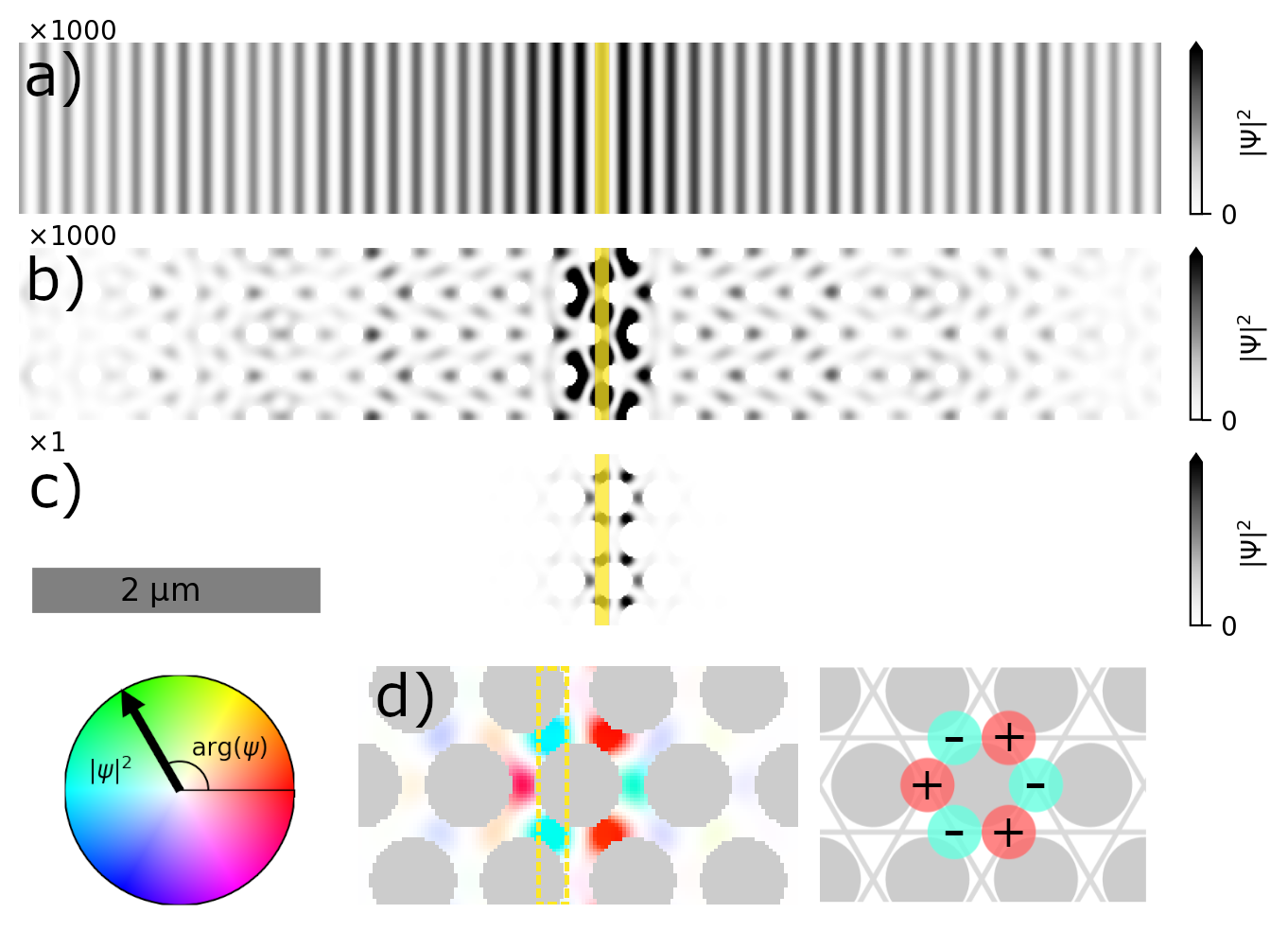}
    \caption{Under periodic boundary conditions, a rectangular region (yellow) is continuously excited at the first flat band frequency of 2.08~GHz. Scaled profiles of $|\Psi|^2$ are plotted after 50~ns for the geometries corresponding to (a,b,c) in Fig.~\ref{fig:Bands}. The strong localization in (c) demonstrates the exceptional flatness of the band. Note the scale differences: the flat band mode is excited to $\approx1000 \times$ the density of its counterparts, making it potentially useful in easily reaching the nonlinear regime. (d) Zoomed-in plot of the mode in (c), with phase encoded in color. (left) Key for color plots and (right) the localized wavefunction of the kagome lattice, which this mode closely resembles. This is an example of how $\Psi$ can be useful to interpret as a wavefunction.}
    \label{fig:Flatband}
\end{figure}

The isotropic flat bands of this system are particularly interesting. A truly flat band is one which has zero group velocity everywhere; its excitations are immobile. Fig.~\ref{fig:Flatband}  illustrates this immobility. A rectangular region is continuously excited at the first flat band frequency. In the non-flat cases [see Figs.~\ref{fig:Flatband}(a) and (b)], the power radiates away as usual. However, the flat band state cannot propagate and is instead excited to $\approx 1000\times$ the density of its counterparts [see Fig.~\ref{fig:Flatband}(c)]. This is a rare feature, making this system potentially useful in nonlinear magnonics experiments (magnon Bose-Einstein condensation, for instance, in which large magnon densities must be achieved) \cite{demokritov_boseeinstein_2006, tiberkevich_excitation_2019}. Outside magnonics, interactions in flat bands on the kagome lattice is a popular topic \cite{wang_quantum_2023} -- given the inherent nonlinearity of magnons, it is possible that this may provide a platform for an analog of interacting phases when this localized state is excited to large magnon number \cite{wu_flat_2007}.

Interestingly, the reason behind localization here can be understood from existing principles in 2D materials. Figure~\ref{fig:Flatband}(d) is a zoomed-in version of Fig.~\ref{fig:Flatband}(c), plotted using a phase-sensitive scheme. This reveals that the excited mode closely resembles the famous localized electronic wavefunction of the kagome lattice, whose localization property can be understood as the result of destructive interference between neighboring sites \cite{rhim_singular_2021, chen_visualizing_2023, multer_imaging_2023}. This spin wave mode is also similar to localized spin excitations observed in the flat bands of layered kagome magnets, which can be understood by the same means \cite{chisnell_topological_2015, riberolles_chiral_2024}.

\begin{figure}
    \centering
    \includegraphics[width=1\linewidth]{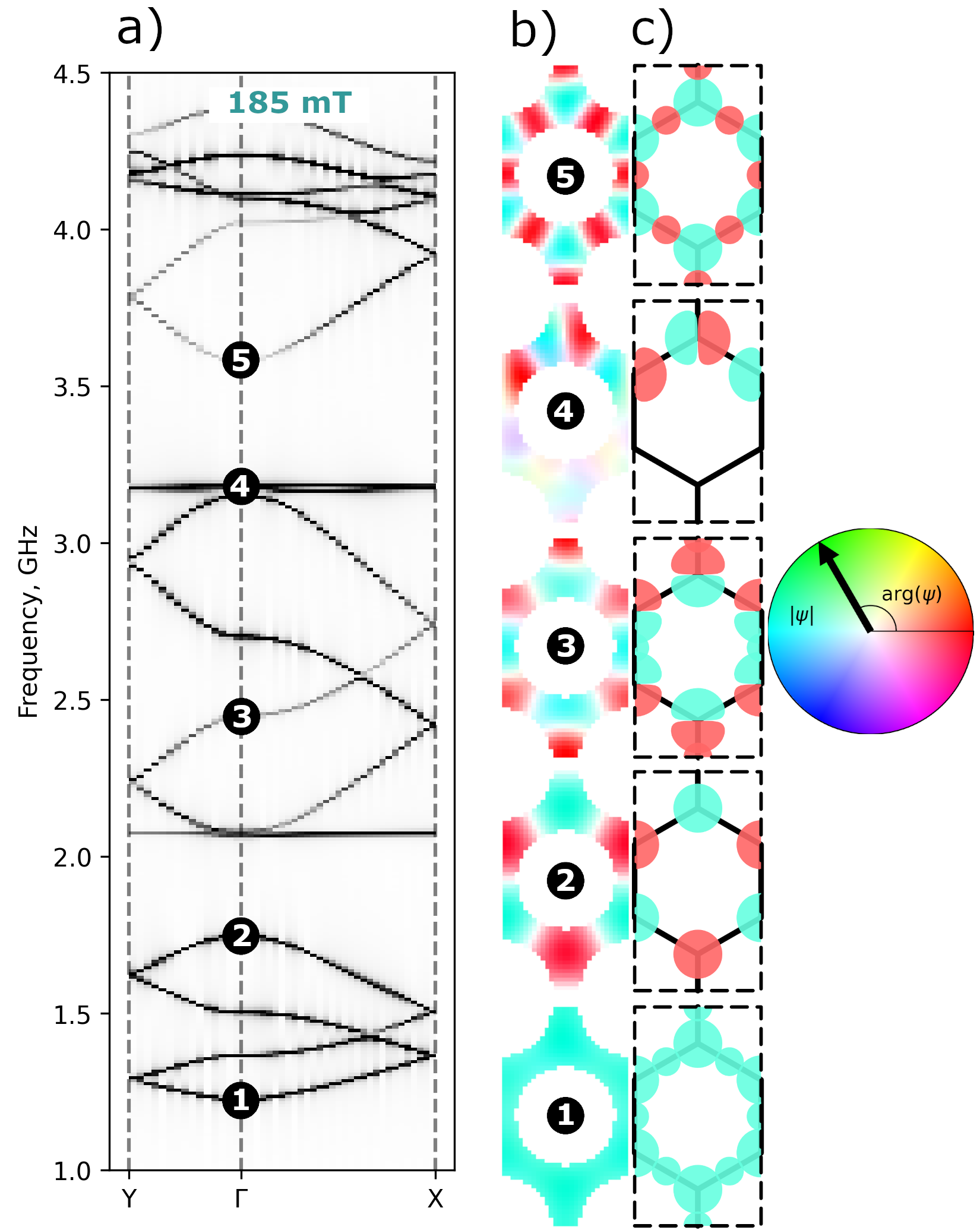}
    \caption{(a) Magnonic band structure for $d/a=0.8$ and $B_{\text{ext}}=185\text{ mT}$ along a few paths in the rectangular BZ. (b) A few examples of Bloch function-like responses at the $\mathbf{\Gamma}$ point, and (c) their reconstructions using a small number of basis 'orbitals,' discussed in more detail in Sec.~\ref{sec:TB_Analysis}.  (right) Key for color plots of $\Psi$. The main finding is that the Bloch functions are simple, suggesting a tight-binding model.}
    \label{fig:Bloch}
\end{figure}

To understand the real nature of the excitations, $\Psi(x,y)$ can be examined at a few $(f,k_x,k_y)$ points. This cell-periodic part of the response can be interpreted as a Bloch function. The magnon band structure is plotted along a few paths in Fig.~\ref{fig:Bloch}(a), and some Bloch-like functions for $\mathbf{\Gamma}$-point modes are plotted in Fig.~\ref{fig:Bloch}(b). One finding is that the lowest set of bands is indeed graphene-like: this figure demonstrates the Dirac point more clearly, but perhaps more incriminating is the fact that $\mathbf{\Gamma}$-point modes $1$ and $2$ in Fig.~\ref{fig:Bloch}(b) are given by acoustic and optical modes of the honeycomb lattice. Actually, there exists a tiny gap at the Dirac point due to the dipole-dipole interaction, which is not easy to resolve (see supplemental Sec.~\ref{Supp-sec:S_DipoleGap} \cite{Supplemental}). This is interesting but will be the primary subject of another study; here, we focus on a simple picture of the band structure.

The most important finding demonstrated by Fig. \ref{fig:Bloch} is that not just the graphene-like modes, but \textit{all} the $\mathbf{\Gamma}$-point modes may be understood with a simple set of basis functions (or "orbitals"). Figure~\ref{fig:Bloch}(c) is an illustration of reconstructions using this set of functions, which consists of: $(1)$ $s$-like spin wave modes on the honeycomb lattice, $(2)$ $p$-like modes on the honeycomb lattice, and $(3)$ $s$-like modes on the kagome lattice, for a total of 9 degrees of freedom per primitive unit cell, or 18 per rectangular cell. 1--2~GHz bands appear to be composed of $s$-like modes and 2--3.5~GHz bands of $p$-like modes -- we refer to these as $s$-bands and $p$-bands, respectively. A natural question is: If the Bloch functions can be reconstructed using these basis orbitals, can the band structure also be reproduced? The latter part of this study is dedicated to this type of understanding and the engineering it enables.

\section{Tight-binding-like Analysis}\label{sec:TB_Analysis}
Taking a step back, a simple Heisenberg spin chain under an externally applied magnetic field has the hamiltonian \cite{stancil_quantum_2009}
\begin{equation}
    \mathcal{H}=-2\dfrac{\mathcal{J}}{\hbar^2}\sum_{\langle ij\rangle}{\mathbf{S}_i \cdot \mathbf{S}_{j}}-\dfrac{g\mu_BB_0}{\hbar}\sum_i{S_{i}^z}
\end{equation}
for an externally applied field $B_0||\hat{z}$ and ferromagnetic exchange parameter $\mathcal{J}$ which couples neighboring spins $\langle ij\rangle$.
    This can be rewritten using spin raising and lowering operators $S^{\pm}\equiv S_x\pm iS_y$. The Holstein-Primakoff (HP) transformation \cite{holstein_field_1940} can then be used to rewrite this problem with magnon operators, assuming total spin $s$ parallel to $\hat{z}$ and small magnon number:
\begin{equation}
\begin{split}
    S^+_i&\approx\hbar\sqrt{2s}a_i \\
    S^-_i&\approx\hbar\sqrt{2s}a_i^\dagger
\end{split}
\end{equation}
\begin{equation}
\begin{split}
    \mathcal{H}=-2\mathcal{J}s\sum_{\langle ij\rangle}{\left( a_i^\dagger a_{j}+a_i a_{j}^\dagger-a_i^\dagger a_i-a_{j}^\dagger a_{j}+s \right)} \\
    +g\mu_BB_0\sum_i{\left(s-a_i^\dagger a_i\right)}
\end{split}
\end{equation}
Defining constants $t=-2\mathcal{J}s$, $\varepsilon=g\mu_BB_0+4\mathcal{J}s$, and keeping only terms with magnon operators, this can be rewritten as
\begin{equation}
    \mathcal{H}=t\sum_{\langle ij\rangle}{\left(a_i^\dagger a_j +a_i a_j^\dagger\right) }+\varepsilon \sum_i{a_i^\dagger a_i}
\end{equation}
This is the generic form of a tight-binding (TB) model for a 1D atomic chain, except that operators refer to magnons in the HP formalism. This suggests that TB models with appropriate parameters can generically describe exchange-only systems with ferromagnetic ground states. The interpretation is the same when these operators act on macrospins in nanostructures \cite{iacocca_reconfigurable_2016}. With this TB picture in mind, it is not a large stretch to expect that nonuniform spin wave modes may also be represented by operators in this model, as long as the hopping parameters $t_{ij}$ and energies $\varepsilon_i$ are properly adjusted. For example, an array of exchange-coupled elements may be described by an effective TB model in which $a_i^\dagger$ creates a magnon in a nonuniform mode $i$, mainly confined to a single magnetic element \cite{shindou_chiral_2013}. To be more precise, the nonuniform mode operators can be thought of as superpositions of magnon operators on many macrospins, with new hopping parameters that are simple to calculate from old parameters. In this picture, the hamiltonian becomes:
\begin{equation}
    \mathcal{H}=\sum_{ij}{t_{ij}\left(a_i^\dagger a_j +a_i a_j^\dagger\right) }+ \sum_i{\varepsilon_i a_i^\dagger a_i}
\end{equation}
in which $i,j$ may now refer to different types of nonuniform modes.

\begin{figure}[b]
    \centering
    \includegraphics[width=1\linewidth]{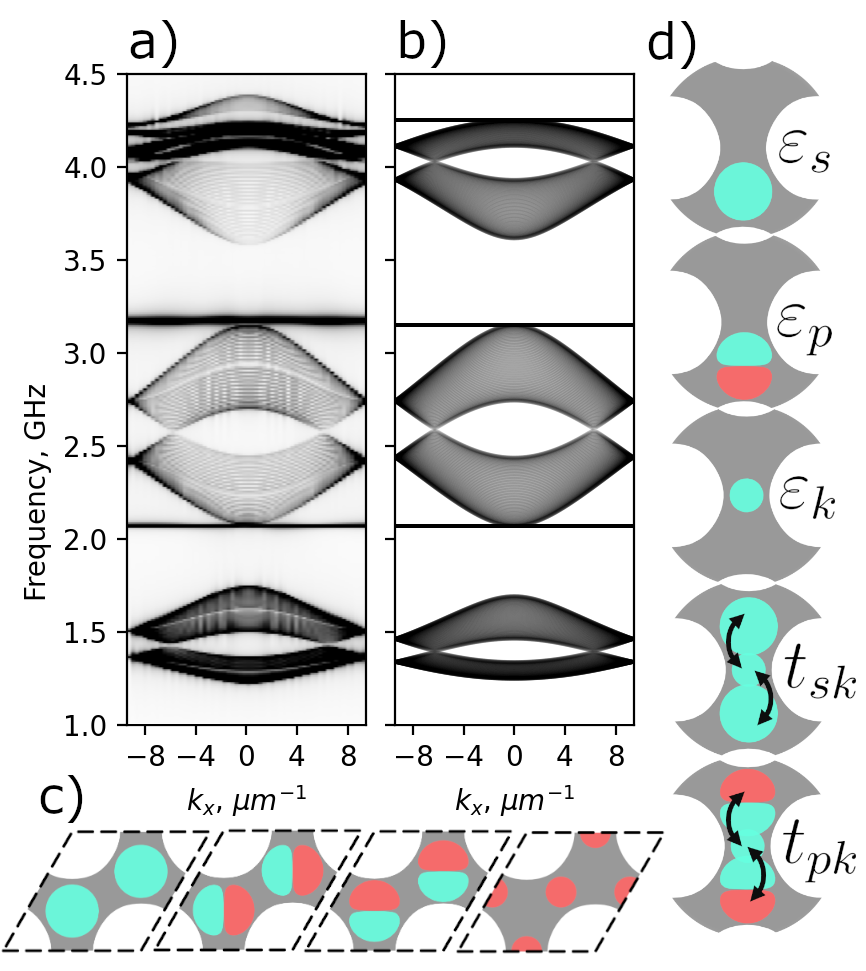}
    \caption{(a) Simulated $k_x$-projected band structure as plotted in Fig.~\ref{fig:Bands}(c), and (b) its 9-band tight-binding model using fitting parameters listed in table \ref{tab:TBParams}. (c) Basis orbitals for the TB model (left to right): $s,p_x,p_y$ orbitals on honeycomb sites, and $s$ orbitals on kagome sites. (d) Illustration of fitting parameters.}
    \label{fig:TB}
\end{figure}

Motivated by the observation that the Bloch functions [see Fig.~\ref{fig:Bloch}(b)] are made of $s$ and $p$ 'orbitals,' we choose this basis for a TB model. Using only nearest-neighbor hoppings, we implement this model using the \textsc{PythTB} package \cite{coh_python_2022} and adjust parameters by hand. Fit parameters are listed in table \ref{tab:TBParams} in units of frequency. Figure~\ref{fig:TB}(a) is the same band structure appearing earlier, and Fig.~\ref{fig:TB}(b) is its corresponding model, displayed in a manner similar to the simulations, using basis orbitals shown in Fig.~\ref{fig:TB}(c) and (d). More precisely, we plot the spectrum of a supercell extended in the $\hat{y}$-direction with semitransparent coloring, and only plot the $k_x$ extent of the rectangular BZ.

\begin{table}
\begin{tabular}{|l|c l |}
    
    \hline
    $\varepsilon_s$ & 1.69 & GHz\\
    $\varepsilon _p$ & 3.15& GHz\\ 
    $\varepsilon _k$ & 3.17& GHz\\
    \hline
    $t_{sk}$ &-0.38 & GHz\\
    $t_{pk}$ & -0.63& GHz\\
    \hline
    
\end{tabular}
\caption{Tight-binding parameters and their values as used in Fig.~\ref{fig:TB}. $\varepsilon_i$ are 'on-site' frequencies for orbital $i$ and $t_{ij}$ are hopping parameters between orbitals $i$ and $j$. Subscripts $s$ and $p$ denote $s$-like and $p_{x,y}$-like modes on the honeycomb lattice. Subscripts $k$ denote $s$-like modes on the kagome lattice. }
\label{tab:TBParams}
\end{table}

It is remarkable that a TB model works well in an anti-dot lattice. The orbitals emerge naturally from the structure without any careful engineering. The fact that a simple model works suggests that this is a general feature of the geometry that has nothing to do with spin waves. Indeed, simply solving the Schr{\"o}dinger equation for a massive particle in a potential resembling this anti-dot lattice can yield a similar band structure (see supplemental Sec. \ref{Supp-sec:S_Schrodinger} \cite{Supplemental}). This phenomenon can be understood from Eq.~\ref{eq:Schrodinger-like}. In fact, as a representation of the simplicity of excitations in this system, similar band structures have been discussed and observed in electronic, photon/polariton, acoustic, and cold-atom systems \cite{milicevic_type-iii_2019, gao_visualization_2024, wu_p_xy-orbital_2008, polini_artificial_2013, mangussi_multi-orbital_2020, ding_experimental_2019}.

Although the model appears to work, there are some inconsistencies. At the $\mathbf{\Gamma}$ point of the lowest band, the dispersion in the simulation is linear due to the dipole-dipole interaction \cite{stancil_magnetostatic_2009}, which is explicitly ignored by the TB model. Small mid-band gaps are also observable in the simulations, but this is only an artifact of the imperfect reconstruction of the hexagonal lattice using pixels, which breaks its $C_6$ symmetry down to $C_2$. Such gaps are not expected for real lattices. In addition, the model is less accurate at high frequencies. This is because the real wavelength of spin waves becomes small compared to the lattice features. For larger hole sizes (smaller lattice features), these high-frequency bands approach a more TB-like form (see supplemental Sec. \ref{Supp-sec:S_MoreBandStructures} \cite{Supplemental}). Other subtle inconsistencies exist, in part due to the TB model's exclusion of the dipole-dipole interaction.

The fact that TB works well is curious but does not teach us anything fundamentally new. However, understanding this band structure in terms of graphene-like $s$ and $p_{x,y}$ excitations enables engineering using existing principles from the field of 2D materials. 

\section{Spin wave engineering}
\subsection{Inversion-broken crystals}\label{sec:Invbroken}
\begin{figure}
    \centering
    \includegraphics[width=1\linewidth]{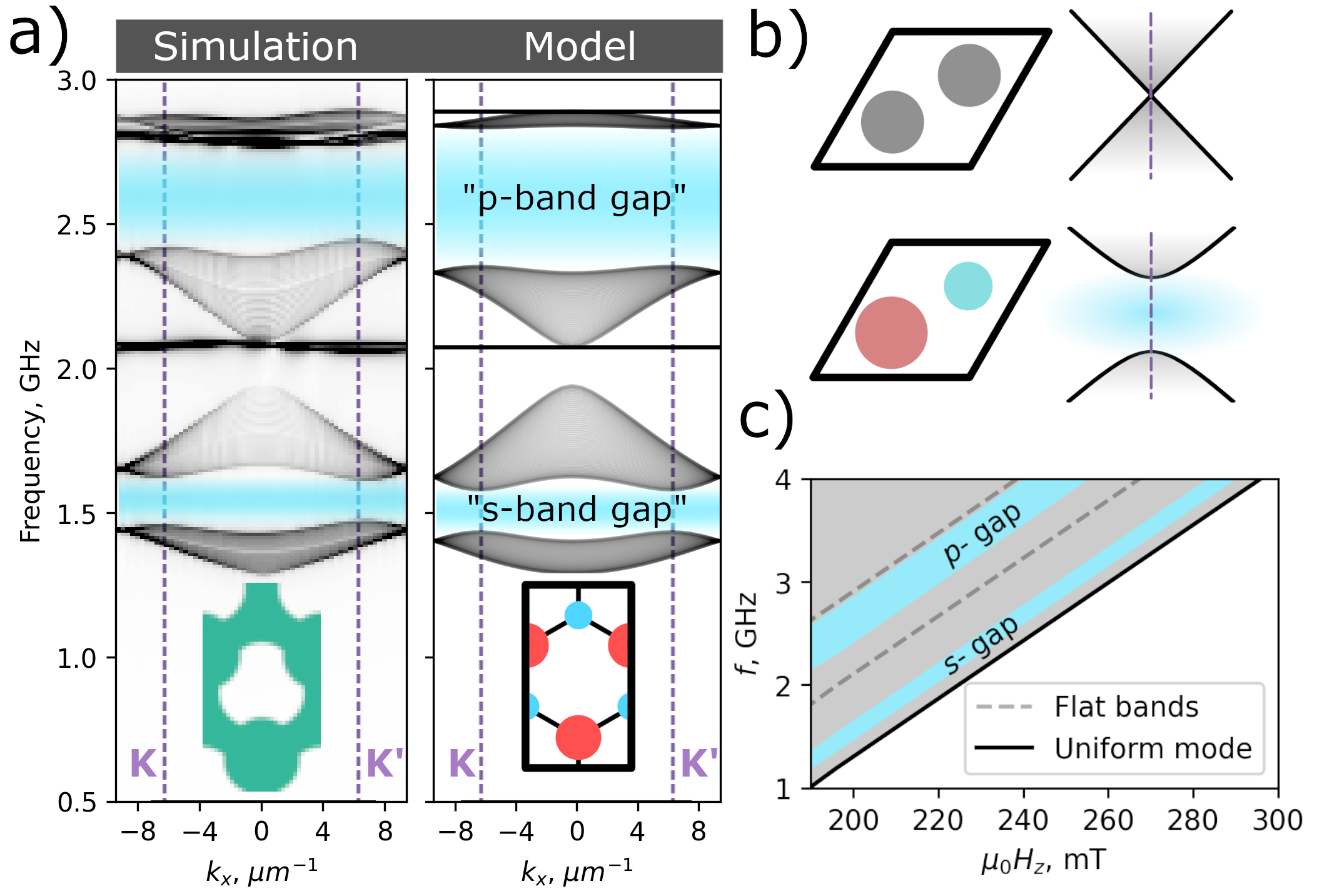}
    \caption{The effect of inversion symmetry breaking on the $s$ and $p$ bands. (a) Band structure for a unit cell with broken symmetry, plotted in the same manner as before. The spectrum can still be modeled with TB. Symmetry breaking enters the model as terms that discriminate between the honeycomb sublattices, $\Delta \varepsilon_{s} $ and $\Delta \varepsilon_{p}$, resulting in band gaps for both the $s$- and $p$-bands. (b) Analogy to graphene and its symmetry-broken sister h-BN: a gap (highlighted in blue) opens at $\mathbf{K}$ and $\mathbf{K'}$ points. Fit parameters are listed in table \ref{tab:TBParams_Invbroken}. (c) The dependence of induced gaps and flat band frequencies on externally applied field, extracted from simulations. This tunability is one of the unique benefits of magnonic systems. }
    \label{fig:InvBreaking}
\end{figure}

In graphene, the masslessness of Dirac electrons is protected in part by inversion symmetry. Hexagonal boron nitride (h-BN), another 2D compound, is isostructural but with different atoms on each honeycomb sublattice. This breaks the inversion symmetry of the unit cell, and the Dirac points gap (for this reason, these symmetry-breaking terms are sometimes called \textit{mass terms}). As a result, h-BN is an insulator. This is shown schematically in Fig.~\ref{fig:InvBreaking}(b). Even stacking graphene on h-BN has this result because of the small amount of symmetry breaking \cite{giovannetti_substrate-induced_2007, xue_scanning_2011}. This principle is powerful because, unless the system is fine-tuned, any method of breaking inversion symmetry should result in a gap.

To mimic the relationship between graphene and h-BN, we create new unit cells that break inversion symmetry and perform simulations again. Figure~\ref{fig:InvBreaking}(a)  displays the resulting band structures in the same manner as before.

As predicted, a gap is present in the inversion-broken geometries. In analogy to h-BN, the symmetry breaking can be described in TB by a difference $\Delta\varepsilon\equiv\frac{1}{2} \left( \varepsilon_B-\varepsilon_A \right)$ in the energies of the two sublattices. For simplicity, this is the only addition made to the model. Controllable spin wave band gaps are desirable for {\em rf} applications; it is interesting to note that this is an example of gaps whose width is tunable using guidance from the analogy to 2D materials. Furthermore, the frequency of these gaps (along with all other features) can be tuned by an externally applied magnetic field as shown in Figure \ref{fig:InvBreaking}(c). This tunability is natural to magnonic crystals and, from a functionality perspective, is a benefit compared to photonic and phononic crystals.
\begin{table}
\begin{tabular}{|l|c l |}
    
    \hline
    $\varepsilon_s$ & 1.93 & GHz\\
    $\varepsilon _p$ & 2.83& GHz\\ 
    $\varepsilon _k$ & 3.28& GHz\\
    \hline
    $\Delta\varepsilon_s$ & 0.12 & GHz\\
    $\Delta\varepsilon_p$ & 0.35 & GHz\\
    \hline
    $t_{sk}$ &-0.45 & GHz\\
    $t_{pk}$ & -0.49& GHz\\
    \hline
    
\end{tabular}
\caption{Tight-binding parameters and their values as used in Fig.~\ref{fig:InvBreaking}, following the same labeling as in table \ref{tab:TBParams}.}
\label{tab:TBParams_Invbroken}
\end{table}

\begin{figure}
    \centering
    \includegraphics[width=1\linewidth]{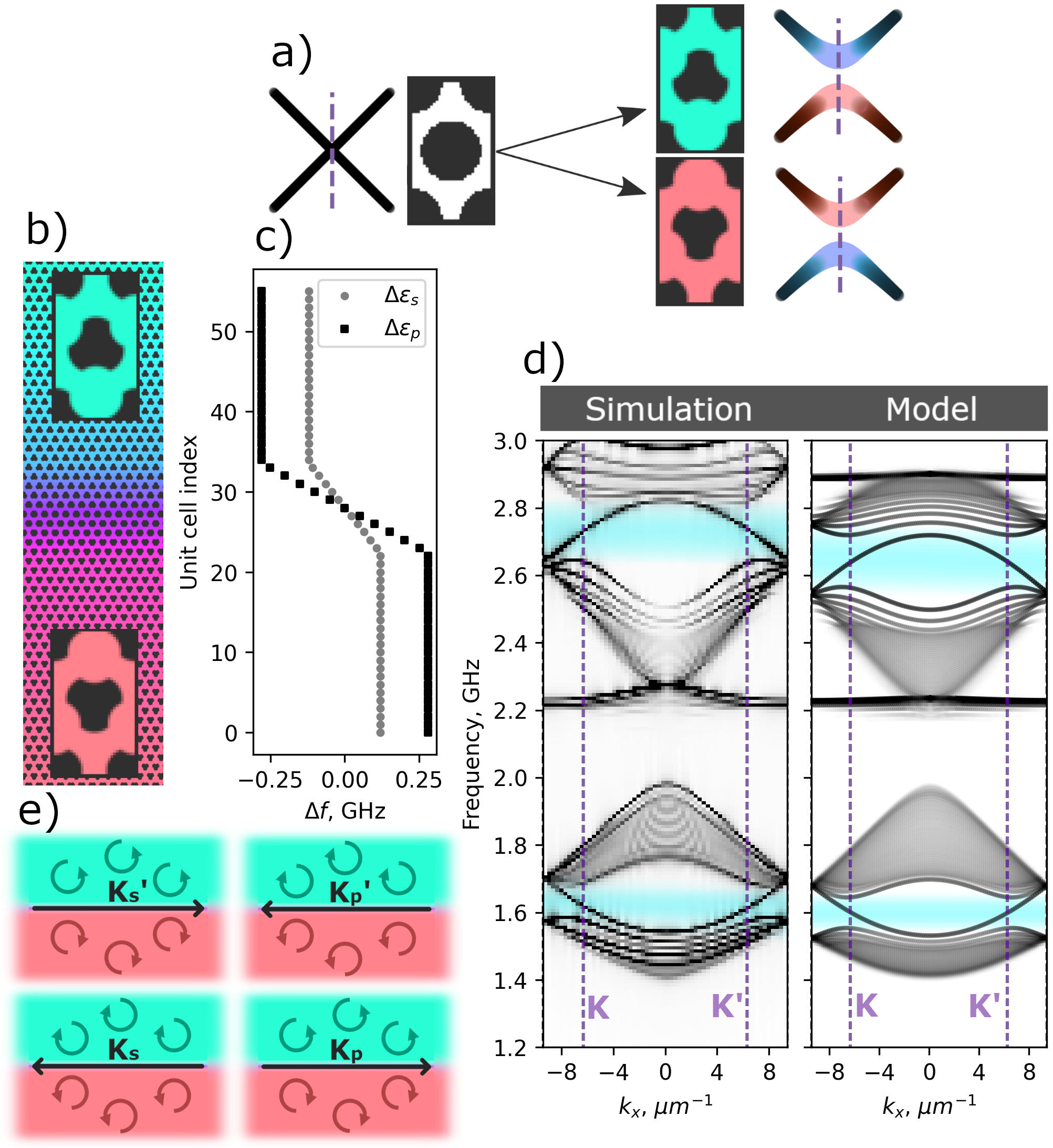}
    \caption{A boundary between different gapped phases. (a) Two distinct gapped phases with different Berry curvature at a given valley. (b) The simulated geometry, including a boundary between the phases. (c) The corresponding TB setup, in which the symmetry-breaking terms change sign along the $\hat{y}$-direction. (d) Simulated and modeled band structures. In the simulation, the excitation is at the boundary. To mimic this, the TB spectrum is plotted with opacity given by amplitude at the boundary. Magnon states bridging the gaps (highlighted in blue) indicates \textit{two instances} of quantum valley-Hall insulator behavior. However, the states bridging the two gaps have opposite propagation: (e) $s$ modes (in the lower gap) have $\mathbf{K'}$-polarized waves moving to the right, while $p$ modes (in the upper gap) have $\mathbf{K'}$-polarized waves moving to the left.}
    \label{fig:Boundary_combined}
\end{figure}

In graphene-like systems, these symmetry-breaking mass terms do more than simply open a gap. For small mass terms, opening such a gap generates sharply peaked Berry curvature near the valleys $\mathbf{K}$ and $\mathbf{K'}$. Integrating the Berry curvature over the vicinity of a single valley yields an approximately quantized value, often referred to as a valley Chern number (VCN) \cite{zhang_valley_2013}. The two valleys carry opposite VCNs, so the total Chern number over the full Brillouin zone is zero. Nevertheless, a change in VCN across a domain wall implies valley-polarized boundary modes \cite{semenoff_domain_2008, jung_valley-hall_2011}, which remain robust as long as intervalley scattering remains weak \cite{zhang_valley_2013}. TB calculations in this $s,p_{x,y}$ system show that symmetry breaking has this effect for \textit{both} the $s$-band gaps and the $p$-band gaps. At a single valley, the sign of the Berry curvature is opposite above and below the gap for both $s$ and $p$ bands, but notably has a different sign in each ($s,p$) case. Upon changing the sign of the symmetry-breaking terms $\Delta\varepsilon_s$ and $\Delta\varepsilon_p$, the sign of curvature flips (see supplemental Sec.~\ref{Supp-sec:S_BerryCurvature} \cite{Supplemental}), schematically illustrated in Fig. ~\ref{fig:Boundary_combined}(a). 

This magnetic thin film geometry, unlike vdW systems, is continuously tunable between distinct gapped phases. Partially inspired by the Jackiw-Rebbi model \cite{jackiw_solitons_1976}, we investigate a boundary between these gapped phases. Figure~\ref{fig:Boundary_combined}(b) shows a gradual phase boundary in which the unit cell changes shape over a few lattice spacings, making a smooth transition between the two phases. 
In the corresponding TB model, we create a long supercell with varying mass terms [Fig.~\ref{fig:Boundary_combined}(c)]. The band structure of the TB model contains the essential features of the simulation, most notably the presence of states that bridge the gap at each valley, highlighted in Fig.~\ref{fig:Boundary_combined}(d). These will be referred to as boundary modes.

In our TB model, as we increase mass terms, Berry curvature is no longer localized to the valleys and the VCN loses its quantization (supplemental Sec. \ref{Supp-sec:S_BerryCurvature} \cite{Supplemental}). This behavior is common in realistic systems \cite{lee_gapped_2022, zhu_design_2018, lee_quantum_2020}, where the absolute value of the VCN is smaller than its quantized value in the small-gap limit and varies smoothly with system parameters. However, since the bulk gap does not close as we vary mass parameters, the boundary modes we observe are adiabatically connected to those in the small-gap limit. They remain (for example, in our simulations) visible well beyond the regime where the VCN is sharply quantized. This phenomenon of boundary-localized states has been seen before in many other graphene-like systems \cite{lu_observation_2017, yang_acoustic_2018, wang_extended_2022, noh_observation_2018, zhang_achieving_2018, wang_ultracompact_2022, funayama_quantum_2024}, but occurs here for both the $s$ and $p$ bands. Because the $\Delta\varepsilon$ terms affect the $s$ and $p$ gaps in different ways, the corresponding magnon boundary modes bridge the gaps with opposite chiralities: in the lower gap, $\mathbf{K}$($\mathbf{K'}$)-valley magnons propagate to the left (right), while in the upper gap, $\mathbf{K}$($\mathbf{K'}$)-valley magnons propagate to the left (right). The direction of this valley-polarized magnon current can be reversed by changing the sign of the symmetry-breaking terms (see supplemental Sec. \ref{Supp-sec:S_InvertedBoundary} \cite{Supplemental}). By calculating the VCNs of the bulk bands in the small-gap limit, we thus identify these boundary states as the magnonic analog of quantum valley Hall (QVH) edge states\cite{yao_edge_2009, zhang_valley_2013}, as illustrated in Fig. ~\ref{fig:Boundary_combined}(e). This is a more accessible realization of the type of state proposed for 2D honeycomb magnets with controllable staggered anisotropies \cite{hidalgo-sacoto_magnon_2020}. For a more complete discussion including Berry curvature maps, see the supplemental Sec. ~\ref{Supp-sec:S_BerryCurvature} \cite{Supplemental}.

\subsection{Topological Magnon Waveguiding}\label{sec:Topological_Waveguiding}
The QVH-like states that are localized to the boundary may be used as a frequency-selective waveguide for valley-polarized magnons.
\begin{figure*}
    \centering
    \includegraphics[width=1\linewidth]{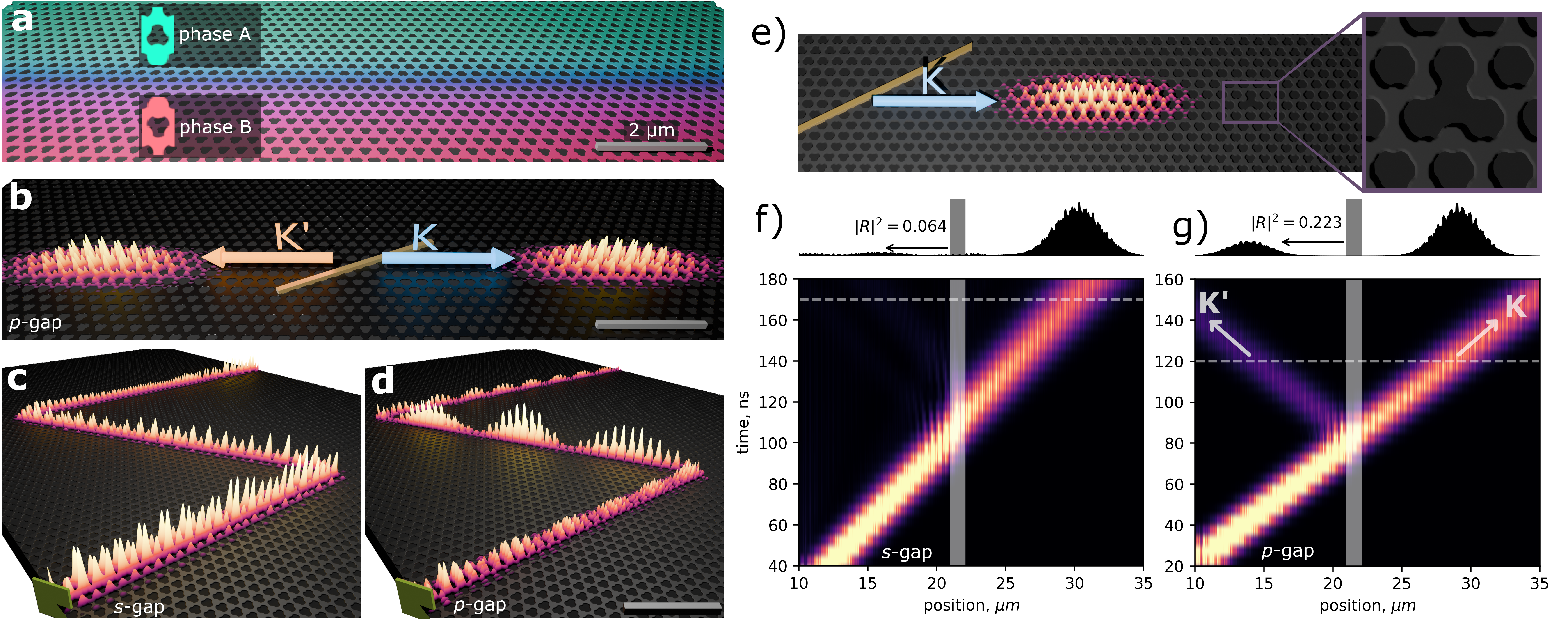}
    \caption{Demonstration of spin wave transport along phase boundaries. (a) Simulation geometry, the same as that in Fig.~\ref{fig:Boundary_combined}.  (b) Spin wave profile $30 \text{ ns}$ after the peak of a $2.7 \text{ GHz}$ pulse (inside the $p$-band gap). Left- and right-moving modes are valley polarized and move only along the boundary. The excitation (yellow) is tilted to match the phase profile of the spin waves and allow efficient coupling. The boundary can be reduced in width and still allow directed transport, even along sharp turns. Profiles are plotted after $500 \text{ ns}$ of continuous excitation at (c) 1.55~GHz in the $s$-band gap, and (d) 2.55~GHz in the $p$-band gap. Unlike the edge states of Chern insulators, because there exist co-localized left- and right- moving edge states, inter-valley scattering is allowed. To demonstrate this, we show (e) Gaussian wavepackets scattering from a point defect formed by removing a honeycomb site. Space-time plots of the spin wave profiles are shown for (f) the $s$-gap mode and (g) the $p$-gap mode with the defect position marked in gray. Profiles are examined after the collision and give calculated reflections of $\approx6\%$ and $\approx22\%$ respectively.}
    \label{fig:3D}
\end{figure*}
In the same geometry as before, shown again in Fig.~\ref{fig:3D}(a), an excitation is applied inside a localized region. At the in-gap frequency of $2.7 \text{ GHz}$, spin waves are launched \textit{only} along the phase boundary, as shown in Fig.~\ref{fig:3D}(b). Because of the valley-specific topological origin of these modes, left- and right- moving modes are oppositely valley polarized -- an interesting feature because valley polarization is occasionally thought of as an information-carrying degree of freedom in discussions of next-generation information technologies \cite{schaibley_valleytronics_2016}. The transition between the insulating phases can be reduced in width, and the states persist, even following sharp turns made by the phase boundary. This phenomenon occurs for both $s$ and $p$ boundary modes [see Figs. \ref{fig:3D}(c) and (d), respectively]. In both cases, magnon transport remains completely localized to the boundary after $500 \text{ ns}$ of continuous excitation.
It is worth noting a few limitations here. Geometry does not only affect the shape of the band structure; there can be a constant frequency offset as a result of the demagnetizing field. This is the same reason $B_{\text{ext}}$ is varied for the different geometries in Fig.~\ref{fig:Bands}. When geometry is varied over space, this varying frequency offset can result in a spectral overlap of the localized mode with other bulk modes. In this case, a large-area excitation like that in Fig.~\ref{fig:3D}(a) would excite both boundary and bulk modes. In other words, the appealing boundary-specific transport demonstrated by Fig.~\ref{fig:3D} may not exist in all geometries with phase boundaries. Also note that these modes are not protected against backscattering, making them fundamentally different from more famous examples of topological edge states in metamaterials, in \cite{shindou_topological_2013,hafezi_imaging_2013} for example. Scattering between valley modes caused by a point defect is demonstrated in Fig. ~\ref{fig:3D}(e-g) for both $s$-gap and $p$-gap modes, returning differing reflection amplitudes for each case. This is discussed in more detail in the supplemental Sec. \ref{Supp-sec:S_Backscattering} \cite{Supplemental}.

\subsection{Isolated defect modes}
\begin{figure}
    \centering
    \includegraphics[width=1\linewidth]{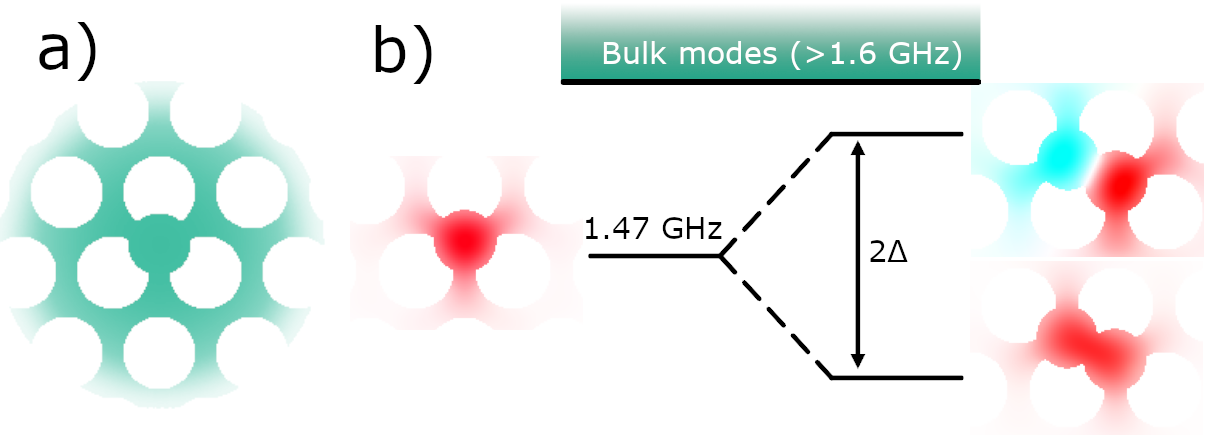}
    \caption{Level structure of point-defect modes under $B_{\text{ext}}=200 \text{ mT } \hat{z}$. (a) Simulation geometry: A single defect resembling that used in the inversion-broken structures of section~\ref{sec:Invbroken}. (b) The localized modes of a single defect versus two adjacent defects. When coupled, they split by $2\Delta\approx0.18\text{ GHz}$, producing a two-level magnon spectrum that appears entirely below the continuum. These modes are plotted in the same fashion as the Bloch-like functions of Fig.~\ref{fig:Bloch}.}
    \label{fig:Defects}
\end{figure}

The unique power of patterned crystals lies in the ability to control its properties in a spatially varying way. In the previous section, this ability was used to exploit valley band topology and waveguide magnons along a one-dimensional boundary. It is also worth discussing magnons that are confined to zero-dimensional point defects. Here we demonstrate a simple case of this: only one honeycomb site is replaced with a disk similar to that in the inversion-broken unit cells [see Fig.~\ref{fig:Defects}(a)], analogous to a substitutional defect in graphene  (nitrogen or boron, for example).
In a phenomenon similar to spin wave "edge modes," \cite{maranville_characterization_2006, mcmichael_edge_2006} A single defect has a localized mode which lies \textit{below} the bulk uniform precession mode. Two such defect modes couple when placed in adjacent sites. Using the TB interpretation, the level should split into symmetric and antisymmetric superpositions of the original modes, separated in frequency by twice the coupling parameter $\Delta$. This is apparently a good interpretation, as evidenced by the mode profiles in Fig.~\ref{fig:Defects}(b). A single-defect mode at $1.47 \text{ GHz}$ splits into a symmetric mode at $1.38 \text{ GHz}$ and an antisymmetric mode at $1.56 \text{ GHz}$ (for details, see supplemental Sec.~\ref{Supp-sec:S_Defects} \cite{Supplemental}). It happens that this whole level structure appears below the continuum, which implies that these defect-localized magnons may be interacted with separately from propagating waves, suggesting further application in magnonics. For the chosen Gilbert damping parameter of $\alpha=10^{-4}$, these modes have quality factors $\approx4100$.

\section{Other TB-like systems}\label{sec:OtherTBSystems}
\begin{figure}[h]
    \centering
    \includegraphics[width=0.8\linewidth]{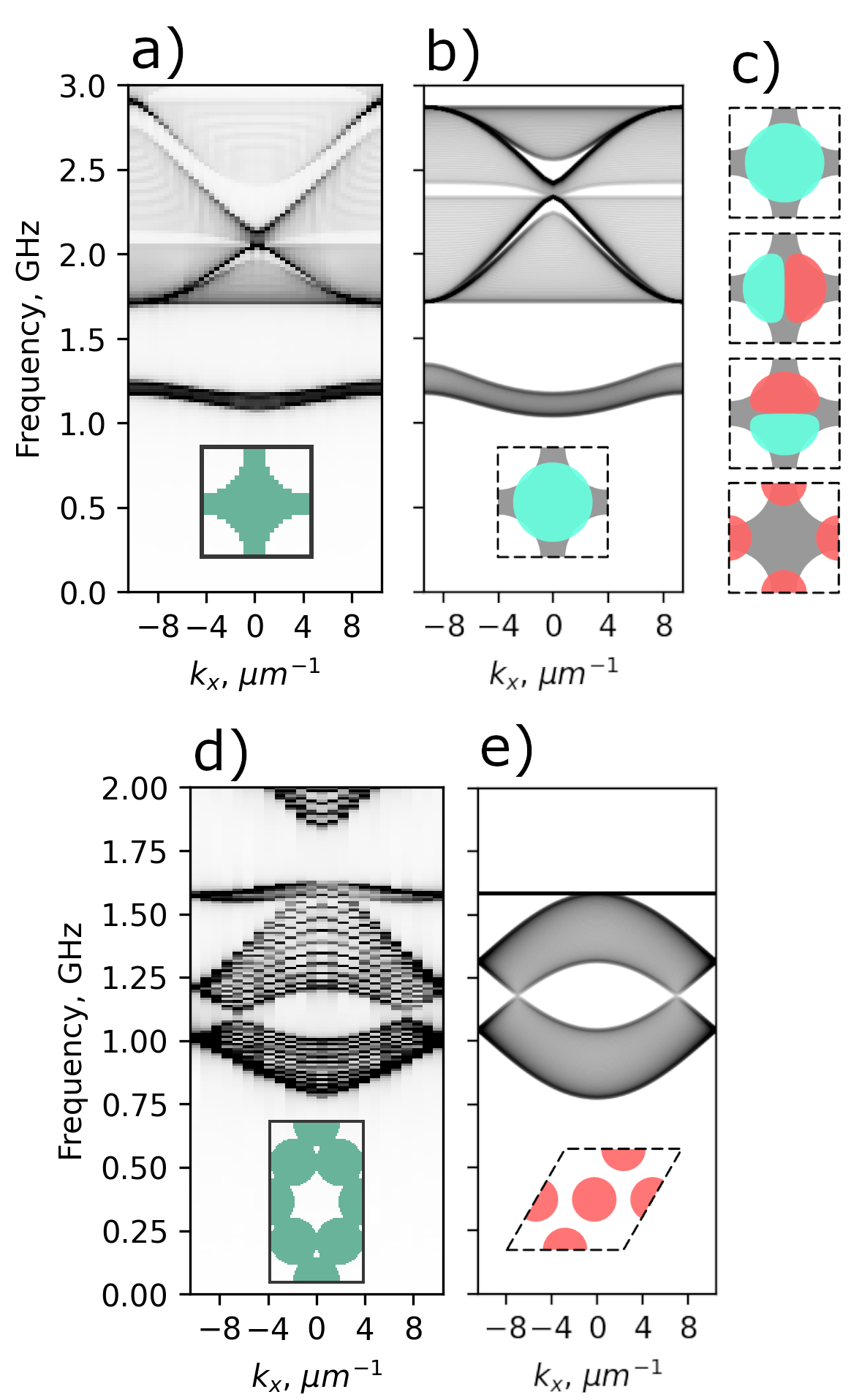}
    \caption{Other geometries which can be thought of as TB-like, using 15-nm YIG films with a lattice spacing a~=~300~nm and $B_{\text{ext}}||\hat{z}=185 \text{ mT}$. (a) The square anti-dot lattice, and (b) its TB model, using (c) basis orbitals similar to those used in the hexagonal anti-dot model. Some inconsistencies are due to hybridization with higher-order modes like $d$-orbitals (see supplemental Sec.~\ref{Supp-sec:S_OtherTB} \cite{Supplemental}), which are not included for simplicity. (d) Shows the artificial kagome lattice and (e) its TB model using only $s$ orbitals on kagome sites.}
    \label{fig:OtherTB}
\end{figure}

In the previous sections, it was demonstrated that a TB-like understanding can be used to engineer magnonic band structures in analogy to a few popular 2D materials. It is worthwhile to show that this TB approach is applicable in at least a few other cases. Fig.~\ref{fig:OtherTB} shows two such examples: a square anti-dot lattice and an artificial kagome lattice.
The magnonic structure of the perpendicularly magnetized square anti-dot lattice [see Fig. \ref{fig:OtherTB}(a)] happens to be analyzable using the same approach. A TB model with $s$ and $p$ orbitals in the middle of the unit cell and $s$ orbitals between these sites can approximate the band structure with the right fitting parameters. Its spectrum is plotted in Fig.~\ref{fig:OtherTB}(b), using the basis orbitals in Fig.~\ref{fig:OtherTB}(c). This square lattice is apparently another case of emergent "atomic orbitals" (see supplemental Sec.~\ref{Supp-sec:S_OtherTB} \cite{Supplemental} for details). There is a simpler (albeit less experimentally feasible) approach to engineer magnonic band structures. Figure~\ref{fig:OtherTB} also shows an artificial kagome lattice, where it is assumed that disks will host $s$-like modes, so the disks are arranged to mimic the kagome lattice [see Fig.~\ref{fig:OtherTB}(d)]. In this case, a simple model of $s$-orbitals on the kagome lattice can approximate the band structure [see Fig. \ref{fig:OtherTB}(e)]. This is similar to results previously reported for magnetic rods embedded in a magnetic matrix \cite{yang_flatbands_2025, liang_dirac_2024}. In this case, the relative decrease in momentum space resolution is due to the fact that the unit cell must be simulated with greater resolution, so the whole simulation field has fewer unit cells for computationally practical reasons.


\section{Experimental feasibility}
The YIG anti-dot lattices studied in this work may be fabricated by standard electron beam lithography procedures including a hard mask and ion etch step, or by focused ion beam (FIB) milling for the fabrication of single devices. There is already a significant amount of work on these structures \cite{neusser_anisotropic_2010, ulrichs_magnonic_2010, bali_high-symmetry_2012, gros_phase_2021, wang_observation_2023}, albeit with somewhat different dimensions. Preliminary work indicates that FIB can yield devices with sufficient scale and tolerances. In the supplemental material Sec.~\ref{Supp-sec:S_Deadlayer} \cite{Supplemental}, we verify that lattices with a thin magnetic dead layer caused by ion damage are expected to yield similar physics. We also verify that in the presence of edge roughness, flat modes, band gaps, defect-localized modes, and QVH-like modes persist with only moderate modification (Sec. ~\ref{Supp-sec:S_Disorder} \cite{Supplemental}). All simulations are conducted for a Gilbert damping parameter $\alpha=10^{-4}$ and $M_{S}=\SI{140}{\kilo  \ampere / \meter}$, which is realistic for YIG \cite{dubs_low_2020,li_epitaxial_2015}. Overall, the anti-dot lattice is a more realistic experimental route to these types of magnonic flat bands compared to existing proposals, which mainly focus on ferromagnetic rods embedded in a matrix\cite{centala_compact_2023, liang_dirac_2024,yang_flatbands_2025}.

\section{Discussion and Conclusion}
Because the magnonic crystals discussed here demonstrate unique properties, it is worth discussing their potential applications.

First, the existence of isotropic flat bands of Fig.~\ref{fig:Flatband}  is notable. In the TB picture, it is trivially easy to achieve flat bands in one sense: an array of decoupled resonators has no dispersion. However, here, the modes are strongly coupled and flatness emerges from the symmetry of the basis orbitals and the geometry of the lattice. Bloch functions in this band contain singularities at band touching points, a property which sometimes awards it the title of "topological flat band" in electronics \cite{liu_orbital_2022, rhim_singular_2021}. Furthermore, these states may be of interest for exploring magnon-magnon interactions, particularly as analogs of interacting electronic phases on the kagome lattice. The localization property of these magnons indicates that they could easily be excited to large magnon numbers, potentially resulting in emergent modes which already have some theoretical description in the hot research area of interacting kagome flat bands. The excitation enhancement due to flat band magnons has been studied before, for example in \cite{wang_broad-wave-vector_2024}.

Second, the topological magnons of Sec.~\ref{sec:Topological_Waveguiding} are, of course, attractive for precise control of magnon propagation. However, they are additionally interesting because we can exercise control over the valley degree of freedom. For example, in the setup of Fig.~\ref{fig:3D}(a), any spin wave device to the right of the antenna should only ever receive $\mathbf{K}$-polarized spin waves at the frequency of $2.7$~GHz. This work may prove an important step in realistic magnon-valleytronics. 

Third, additional control over the valley degree of freedom could reasonably be exercised here through an inhomogeneous pattern that mimics a strain field. Inhomogeneous strain fields can result in an effective gauge field for Dirac point excitations, commonly called a pseudo-magnetic field \cite{pereira_strain_2009, kang_pseudo-magnetic_2021}. The freedom associated with patterning in this system might be used to realize magnon valley filters, lenses, collimators, and (pseudo-)Landau levels for magnons \cite{sun_magnon_2021, wei_strain-engineered_2024, jamadi_direct_2020}.

Fourth, point defects such as those in Fig.~\ref{fig:Defects} may be useful in magnonics because of their spatial localization and spectral isolation. This implies that a carefully placed waveguide could selectively excite these quasi-uniform localized modes -- making them potentially useful as a magnonic memory because spin waves will not radiate away from the defect. Additionally, the fact that a coupled two-mode structure of such defects still lies below the continuum is interesting. We envision classical spin wave analogs of effects like induced transparency and Rabi oscillation. If sufficiently long magnon lifetimes can be achieved, possibly aided by the localization of the modes\cite{serha_ultra-long-living_2025}, then this may also be an interesting perspective for designing quantum magnonic systems.

Fifth, these findings are generalizable in a few ways. Sec.~\ref{sec:OtherTBSystems} demonstrates that tight binding is somewhat accurate in a few other cases of perpendicularly magnetized YIG films, implying that the 2D-analog engineering done in the previous sections can be extended to other magnon systems. However, in Sec. \ref{sec:TB_Analysis} we mention that both tight-binding and the Schr{\"o}dinger equation can reproduce band structures similar to the ones observed here -- implying that some of these engineering strategies are applicable not only to magnonic systems, but any system which can reasonably be described by a simple Schr{\"o}dinger equation.

Finally, because there exists similar physics in other analog systems \cite{milicevic_type-iii_2019, gao_visualization_2024, wu_p_xy-orbital_2008, polini_artificial_2013, mangussi_multi-orbital_2020, ding_experimental_2019}, it is worth noting a few novelties and benefits associated with this magnonic system. The literature tends to focus on $s$-like modes, but we show that it's also possible to accomplish valley-specific physics with $p$-like modes, and we show that it can be understood through the same TB lens. This amplifies the potential for valley-tronics because there are two channels through which valley information may propagate. There are also benefits that are unique to magnonic systems. Most importantly, the whole band structure is tunable not just by geometry, but also by externally applied field, as was demonstrated in Fig. ~\ref{fig:InvBreaking}(c). This suggests that a constant microwave signal may be tuned on the fly to be blocked by band gaps, waveguided, valley- or orbitally polarized, strongly localized in flat bands, interacting with defect modes, etc. simply by varying the external field. Spin waves also are associated with a unique combination of small wavelengths and low frequencies; it is a significant result that a microwave signal can be localized to these phase boundaries by some band topology manipulation.

In conclusion, patterning a hexagonal array of holes into perpendicularly magnetized YIG thin films is found to lead to some unexpected features. Namely, the band structure mimics a few unique properties of 2D systems, and can be understood using a tight-binding model. This new understanding allows the engineering of magnonic states using existing principles from 2D materials to produce controllable band gaps, topological magnons along 1D channels, and spectrally isolated magnons at 0D point defects. These design principles seem to be generalizable to other geometries and even other excitations, but here represent an important step in achieving and manipulating physical states that are usually reserved for van der Waals systems.

\section{Acknowledgements}
All aspects of this work were mainly supported by the Illinois Materials Research Science and Engineering Center (MRSEC) under grant no. DMR-2309037. Additional contributions to data analysis and manuscript preparation were supported by the U.S. DOE, Office of Science, Basic Energy Sciences, Materials Science and Engineering Division under contract No. DE-SC0022060. Other contributions to analysis were supported by the U.S. Office of Naval Research (ONR) Multidisciplinary University Research Initiative (MURI) Grant N00014-20-1-2325 on Robust Photonic Materials with High-Order Topological Protection for support. We thank Taylor Hughes for stimulating and fruitful discussions.
\section{Suppelemental Material and Data Availability}
See Supplemental Material \cite{Supplemental} for a document containing items cited in the text. Codes used for simulation and analysis are also openly available in Ref. ~\cite{sim_code}.

\section{Appendix}
\subsection{Schr{\"o}dinger-like form of the LL Equation}\label{sec:App_Schrodinger}
Beginning with the Landau-Lifshitz-Gilbert equation:
\begin{equation*}
\partial_t\mathbf{M}=-\gamma\mu_0\mathbf{M}\times\mathbf{H}_{\text{eff}} + \dfrac{\alpha}
{M_S}\left(\mathbf{M}\times\partial_t\mathbf{M}\right)
\end{equation*}for magnetization $\mathbf{M}$ of magnitude $M_S$, gyromagnetic ratio $\gamma$ and permeability of free space $\mu_0$. Neglecting damping and rewriting in terms of the reduced magnetization vector $\mathbf{m}\equiv\mathbf{ M}/M_S$,
\begin{equation*}
    \partial_t \mathbf{m}=-\gamma\mu_0\mathbf{m}\times\mathbf{H}_{\text{eff}}
\end{equation*}
Assuming $\mathbf{H}_{\text{eff}}||\hat{z}\equiv H$, $\mathbf{m}\approx \hat{z}$, the linear limit,
\begin{align*}
\qquad\partial_t m_x&=-\gamma \mu_0H m_y\\
\qquad \partial_t m_y&=\gamma \mu_0H m_x\\
\partial_t(m_x+im_y)&=\gamma \mu_0H(im_x-m_y)\\\\
\text{defining:}\qquad \Psi &\equiv m_x+im_y\\\\
\partial_t \Psi&=i\gamma \mu_0H \Psi\\
\end{align*}
Because we have introduced no $\mathbf{H}_{\text{eff}}$ term which couples spins over space, this should also be true for a set of macrospins organized in space: 
\begin{equation*}
    \partial_t\Psi(\mathbf{r})=i\gamma  \mu_0H(\mathbf{r}) \Psi(\mathbf{r})
\end{equation*}
Including the exchange interaction of strength $D_{\text{ex}}$ in the effective field (and assuming its direction is $+\hat{z}$), this can be written in a form \cite{landau_magnetism_nodate, herring_theory_1951, lim_ferromagnetic_2021}:
\begin{equation*}
    \partial_t\Psi(\mathbf{r})=i\gamma  \mu_0\left(H(\mathbf{r})-D_{\text{ex}}\nabla^2\right) \Psi(\mathbf{r})
\end{equation*}
Which resembles a Schr{\"o}dinger equation for a massive particle in a potential $\gamma\mu_0H(\mathbf{r})$. See ref. \cite{lim_ferromagnetic_2021} for an application of this idea to quantized spin waves modes in finite-size systems.
\bibliography{references-2}

@article{novoselov_electric_2004,
	title = {Electric {Field} {Effect} in {Atomically} {Thin} {Carbon} {Films}},
	volume = {306},
	url = {https://www.science.org/doi/abs/10.1126/science.1102896},
	doi = {10.1126/science.1102896},
	number = {5696},
	urldate = {2025-05-22},
	journal = {Science},
	author = {Novoselov, K. S. and Geim, A. K. and Morozov, S. V. and Jiang, D. and Zhang, Y. and Dubonos, S. V. and Grigorieva, I. V. and Firsov, A. A.},
	month = oct,
	year = {2004},
	pages = {666--669},
}

@article{katsnelson_chiral_2006,
	title = {Chiral tunnelling and the {Klein} paradox in graphene},
	volume = {2},
	copyright = {http://www.springer.com/tdm},
	issn = {1745-2473, 1745-2481},
	url = {https://www.nature.com/articles/nphys384},
	doi = {10.1038/nphys384},
	language = {english},
	number = {9},
	urldate = {2025-05-23},
	journal = {Nat. Phys.},
	author = {Katsnelson, M. I. and Novoselov, K. S. and Geim, A. K.},
	month = sep,
	year = {2006},
	pages = {620--625},
}

@article{guinea_energy_2010,
	title = {Energy gaps and a zero-field quantum {Hall} effect in graphene by strain engineering},
	volume = {6},
	copyright = {https://www.springernature.com/gp/researchers/text-and-data-mining},
	issn = {1745-2473, 1745-2481},
	url = {https://www.nature.com/articles/nphys1420},
	doi = {10.1038/nphys1420},
	language = {english},
	number = {1},
	urldate = {2025-05-23},
	journal = {Nat. Phys.},
	author = {Guinea, F. and Katsnelson, M. I. and Geim, A. K.},
	month = jan,
	year = {2010},
	
	pages = {30--33},
}

@article{li_observation_2007,
	title = {Observation of {Landau} levels of {Dirac} fermions in graphite},
	volume = {3},
	copyright = {https://www.springernature.com/gp/researchers/text-and-data-mining},
	issn = {1745-2473, 1745-2481},
	url = {https://www.nature.com/articles/nphys653},
	doi = {10.1038/nphys653},
	language = {english},
	number = {9},
	urldate = {2025-05-23},
	journal = {Nat. Phys.},
	author = {Li, Guohong and Andrei, Eva Y.},
	month = sep,
	year = {2007},
	pages = {623--627},
}

@article{podzorov_high-mobility_2004,
	title = {High-mobility field-effect transistors based on transition metal dichalcogenides},
	volume = {84},
	issn = {0003-6951, 1077-3118},
	url = {https://pubs.aip.org/apl/article/84/17/3301/508351/High-mobility-field-effect-transistors-based-on},
	doi = {10.1063/1.1723695},
	language = {english},
	number = {17},
	urldate = {2025-05-23},
	journal = {Appl. Phys Lett.},
	author = {Podzorov, V. and Gershenson, M. E. and Kloc, Ch. and Zeis, R. and Bucher, E.},
	month = apr,
	year = {2004},
	pages = {3301--3303},
}

@article{novoselov_two-dimensional_2005,
	title = {Two-dimensional atomic crystals},
	volume = {102},
	url = {https://www.pnas.org/doi/full/10.1073/pnas.0502848102},
	doi = {10.1073/pnas.0502848102},
	number = {30},
	urldate = {2025-05-23},
	journal = {Proc. Natl. Acad. Sci. U.S.A.},
	author = {Novoselov, K. S. and Jiang, D. and Schedin, F. and Booth, T. J. and Khotkevich, V. V. and Morozov, S. V. and Geim, A. K.},
	month = jul,
	year = {2005},
	pages = {10451--10453},
}

@article{ugeda_characterization_2016,
	title = {Characterization of collective ground states in single-layer {NbSe$_2$}},
	volume = {12},
	copyright = {2015 Springer Nature Limited},
	issn = {1745-2481},
	url = {https://www.nature.com/articles/nphys3527},
	doi = {10.1038/nphys3527},
	language = {english},
	number = {1},
	urldate = {2025-05-23},
	journal = {Nat. Phys.},
	author = {Ugeda, Miguel M. and Bradley, Aaron J. and Zhang, Yi and Onishi, Seita and Chen, Yi and Ruan, Wei and Ojeda-Aristizabal, Claudia and Ryu, Hyejin and Edmonds, Mark T. and Tsai, Hsin-Zon and Riss, Alexander and Mo, Sung-Kwan and Lee, Dunghai and Zettl, Alex and Hussain, Zahid and Shen, Zhi-Xun and Crommie, Michael F.},
	month = jan,
	year = {2016},
	pages = {92--97},
}

@article{cao_unconventional_2018,
	title = {Unconventional superconductivity in magic-angle graphene superlattices},
	volume = {556},
	copyright = {2018 Macmillan Publishers Limited, part of Springer Nature. All rights reserved.},
	issn = {1476-4687},
	url = {https://www.nature.com/articles/nature26160},
	doi = {10.1038/nature26160},
	language = {english},
	number = {7699},
	urldate = {2025-05-23},
	journal = {Nature},
	author = {Cao, Yuan and Fatemi, Valla and Fang, Shiang and Watanabe, Kenji and Taniguchi, Takashi and Kaxiras, Efthimios and Jarillo-Herrero, Pablo},
	month = apr,
	year = {2018},
	pages = {43--50},
}

@article{zeng_thermodynamic_2023,
	title = {Thermodynamic evidence of fractional {Chern} insulator in moiré {MoTe$_2$}},
	volume = {622},
	copyright = {2023 The Author(s), under exclusive licence to Springer Nature Limited},
	issn = {1476-4687},
	url = {https://www.nature.com/articles/s41586-023-06452-3},
	doi = {10.1038/s41586-023-06452-3},
	language = {english},
	number = {7981},
	urldate = {2025-05-23},
	journal = {Nature},
	author = {Zeng, Yihang and Xia, Zhengchao and Kang, Kaifei and Zhu, Jiacheng and Knüppel, Patrick and Vaswani, Chirag and Watanabe, Kenji and Taniguchi, Takashi and Mak, Kin Fai and Shan, Jie},
	month = oct,
	year = {2023},
	pages = {69--73},
}

@article{gong_discovery_2017,
	title = {Discovery of intrinsic ferromagnetism in two-dimensional van der {Waals} crystals},
	volume = {546},
	copyright = {2017 Macmillan Publishers Limited, part of Springer Nature. All rights reserved.},
	issn = {1476-4687},
	url = {https://www.nature.com/articles/nature22060},
	doi = {10.1038/nature22060},
	language = {english},
	number = {7657},
	urldate = {2025-05-23},
	journal = {Nature},
	author = {Gong, Cheng and Li, Lin and Li, Zhenglu and Ji, Huiwen and Stern, Alex and Xia, Yang and Cao, Ting and Bao, Wei and Wang, Chenzhe and Wang, Yuan and Qiu, Z. Q. and Cava, R. J. and Louie, Steven G. and Xia, Jing and Zhang, Xiang},
	month = jun,
	year = {2017},
	pages = {265--269},
}

@article{huang_layer-dependent_2017,
	title = {Layer-dependent ferromagnetism in a van der {Waals} crystal down to the monolayer limit},
	volume = {546},
	copyright = {2017 Macmillan Publishers Limited, part of Springer Nature. All rights reserved.},
	issn = {1476-4687},
	url = {https://www.nature.com/articles/nature22391},
	doi = {10.1038/nature22391},
	language = {english},
	number = {7657},
	urldate = {2025-05-23},
	journal = {Nature},
	author = {Huang, Bevin and Clark, Genevieve and Navarro-Moratalla, Efrén and Klein, Dahlia R. and Cheng, Ran and Seyler, Kyle L. and Zhong, Ding and Schmidgall, Emma and McGuire, Michael A. and Cobden, David H. and Yao, Wang and Xiao, Di and Jarillo-Herrero, Pablo and Xu, Xiaodong},
	month = jun,
	year = {2017},
	pages = {270--273},
}

@article{han_graphene_2014,
	title = {Graphene spintronics},
	volume = {9},
	copyright = {2014 Springer Nature Limited},
	issn = {1748-3395},
	url = {https://www.nature.com/articles/nnano.2014.214},
	doi = {10.1038/nnano.2014.214},
	language = {english},
	number = {10},
	urldate = {2025-05-23},
	journal = {Nat. Nanotechnol.},
	author = {Han, Wei and Kawakami, Roland K. and Gmitra, Martin and Fabian, Jaroslav},
	month = oct,
	year = {2014},
	pages = {794--807},
}

@article{wu_p_xy-orbital_2008,
	title = {$p_{x,y}$-orbital counterpart of graphene: {Cold} atoms in the honeycomb optical lattice},
	volume = {77},
	shorttitle = {$\{p\}\_\{x,y\}$-orbital counterpart of graphene},
	url = {https://link.aps.org/doi/10.1103/PhysRevB.77.235107},
	doi = {10.1103/PhysRevB.77.235107},
	number = {23},
	urldate = {2025-05-23},
	journal = {Phys. Rev. B},
	author = {Wu, Congjun and Das Sarma, S.},
	month = jun,
	year = {2008},
	pages = {235107},
}

@article{liu_thickness-dependent_2019,
	title = {Thickness-dependent magnetic order in {CrI$_3$} single crystals},
	volume = {9},
	copyright = {2019 The Author(s)},
	issn = {2045-2322},
	url = {https://www.nature.com/articles/s41598-019-50000-x},
	doi = {10.1038/s41598-019-50000-x},
	language = {english},
	number = {1},
	urldate = {2025-05-23},
	journal = {Sci. Rep.},
	author = {Liu, Yu and Wu, Lijun and Tong, Xiao and Li, Jun and Tao, Jing and Zhu, Yimei and Petrovic, C.},
	month = sep,
	year = {2019},
	pages = {13599},
}

@article{bedoya-pinto_intrinsic_2021,
	title = {Intrinsic {2D}-{XY} ferromagnetism in a van der {Waals} monolayer},
	volume = {374},
	url = {https://www.science.org/doi/10.1126/science.abd5146},
	doi = {10.1126/science.abd5146},
	number = {6567},
	urldate = {2025-05-23},
	journal = {Science},
	author = {Bedoya-Pinto, Amilcar and Ji, Jing-Rong and Pandeya, Avanindra K. and Gargiani, Pierluigi and Valvidares, Manuel and Sessi, Paolo and Taylor, James M. and Radu, Florin and Chang, Kai and Parkin, Stuart S. P.},
	month = oct,
	year = {2021},
	pages = {616--620},
}

@article{liu_vapor_2020,
	title = {Vapor {Deposition} of {Magnetic} {Van} der {Waals} {NiI$_2$} {Crystals}},
	volume = {14},
	issn = {1936-0851},
	url = {https://doi.org/10.1021/acsnano.0c04499},
	doi = {10.1021/acsnano.0c04499},
	number = {8},
	urldate = {2025-05-23},
	journal = {ACS Nano},
	author = {Liu, Haining and Wang, Xinsheng and Wu, Juanxia and Chen, Yuansha and Wan, Jing and Wen, Rui and Yang, Jinbo and Liu, Ying and Song, Zhigang and Xie, Liming},
	month = aug,
	year = {2020},
	pages = {10544--10551},
}

@article{bikaljevic_noncollinear_2021,
	title = {Noncollinear {Magnetic} {Order} in {Two}-{Dimensional} {NiBr$_2$} {Films} {Grown} on {Au}(111)},
	volume = {15},
	issn = {1936-0851},
	url = {https://doi.org/10.1021/acsnano.1c05221},
	doi = {10.1021/acsnano.1c05221},
	number = {9},
	urldate = {2025-05-23},
	journal = {ACS Nano},
	author = {Bikaljevi{\'c}, Djuro and Gonz{\'a}lez-Orellana, Carmen and Pe{\~n}a-D{\'i}az, Marina and Steiner, Dominik and Dreiser, Jan and Gargiani, Pierluigi and Foerster, Michael and Ni{\~n}o, Miguel Angel and Aballe, Luc{\'i}a and Ruiz-Gomez, Sandra and Friedrich, Niklas and Hieulle, Jeremy and Jingcheng, Li and Ilyn, Maxim and Rogero, Celia and Pascual, Jos{\'e} Ignacio},
	month = sep,
	year = {2021},
	pages = {14985--14995},
}

@article{kruglyak_magnonics_2010,
	title = {Magnonics},
	volume = {43},
	issn = {0022-3727},
	url = {https://dx.doi.org/10.1088/0022-3727/43/26/264001},
	doi = {10.1088/0022-3727/43/26/264001},
	language = {english},
	number = {26},
	urldate = {2025-05-23},
	journal = {J. Phys. D: Appl. Phys.},
	author = {Kruglyak, V V and Demokritov, S O and Grundler, D},
	month = jun,
	year = {2010},
	pages = {264001},
}

@article{serga_yig_2010,
	title = {{YIG} magnonics},
	volume = {43},
	issn = {0022-3727},
	url = {https://dx.doi.org/10.1088/0022-3727/43/26/264002},
	doi = {10.1088/0022-3727/43/26/264002},
	language = {english},
	number = {26},
	urldate = {2025-05-23},
	journal = {J. Phys. D: Appl. Phys.},
	author = {Serga, A A and Chumak, A V and Hillebrands, B},
	month = jun,
	year = {2010},
	pages = {264002},
}

@article{chumak_magnon_2015,
	title = {Magnon spintronics},
	volume = {11},
	copyright = {2014 Springer Nature Limited},
	issn = {1745-2481},
	url = {https://www.nature.com/articles/nphys3347},
	doi = {10.1038/nphys3347},
	language = {english},
	number = {6},
	urldate = {2025-05-23},
	journal = {Nat. Phys.},
	author = {Chumak, A. V. and Vasyuchka, V. I. and Serga, A. A. and Hillebrands, B.},
	month = jun,
	year = {2015},
	pages = {453--461},
}

@article{wang_inverse-design_2021,
	title = {Inverse-design magnonic devices},
	volume = {12},
	copyright = {2021 The Author(s)},
	issn = {2041-1723},
	url = {https://www.nature.com/articles/s41467-021-22897-4},
	doi = {10.1038/s41467-021-22897-4},
	language = {english},
	number = {1},
	urldate = {2025-05-27},
	journal = {Nat. Commun.},
	author = {Wang, Qi and Chumak, Andrii V. and Pirro, Philipp},
	month = may,
	year = {2021},
	pages = {2636},
}

@article{zenbaa_universal_2025,
	title = {A universal inverse-design magnonic device},
	volume = {8},
	copyright = {2025 The Author(s), under exclusive licence to Springer Nature Limited},
	issn = {2520-1131},
	url = {https://www.nature.com/articles/s41928-024-01333-7},
	doi = {10.1038/s41928-024-01333-7},
	language = {english},
	number = {2},
	urldate = {2025-05-27},
	journal = {Nat. Electron.},
	author = {Zenbaa, Noura and Abert, Claas and Majcen, Fabian and Kerber, Michael and Serha, Rostyslav O. and Knauer, Sebastian and Wang, Qi and Schrefl, Thomas and Suess, Dieter and Chumak, Andrii V.},
	month = feb,
	year = {2025},
	pages = {106--115},
}

@article{korber_pattern_2023,
	title = {Pattern recognition in reciprocal space with a magnon-scattering reservoir},
	volume = {14},
	copyright = {2023 The Author(s)},
	issn = {2041-1723},
	url = {https://www.nature.com/articles/s41467-023-39452-y},
	doi = {10.1038/s41467-023-39452-y},
	language = {english},
	number = {1},
	urldate = {2025-05-27},
	journal = {Nat. Commun.},
	author = {Körber, Lukas and Heins, Christopher and Hula, Tobias and Kim, Joo-Von and Thlang, Sonia and Schultheiss, Helmut and Fassbender, Jürgen and Schultheiss, Katrin},
	month = jul,
	year = {2023},
	pages = {3954},
}

@article{papp_nanoscale_2021,
	title = {Nanoscale neural network using non-linear spin-wave interference},
	volume = {12},
	copyright = {2021 The Author(s)},
	issn = {2041-1723},
	url = {https://www.nature.com/articles/s41467-021-26711-z},
	doi = {10.1038/s41467-021-26711-z},
	language = {english},
	number = {1},
	urldate = {2025-05-27},
	journal = {Nat. Commun.},
	author = {Papp, Adam and Porod, Wolfgang and Csaba, Gyorgy},
	month = nov,
	year = {2021},
	pages = {6422},
}

@article{schneeloch_gapless_2022,
	title = {Gapless {Dirac} magnons in {CrCl$_3$}},
	volume = {7},
	copyright = {2022 The Author(s)},
	issn = {2397-4648},
	url = {https://www.nature.com/articles/s41535-022-00473-3},
	doi = {10.1038/s41535-022-00473-3},
	language = {english},
	number = {1},
	urldate = {2025-05-27},
	journal = {npc Quantum Mater.},
	author = {Schneeloch, John A. and Tao, Yu and Cheng, Yongqiang and Daemen, Luke and Xu, Guangyong and Zhang, Qiang and Louca, Despina},
	month = jun,
	year = {2022},
	pages = {1--7},
}

@article{chen_topological_2018,
	title = {Topological {Spin} {Excitations} in {Honeycomb} {Ferromagnet} {CrI$_3$}},
	volume = {8},
	url = {https://link.aps.org/doi/10.1103/PhysRevX.8.041028},
	doi = {10.1103/PhysRevX.8.041028},
	number = {4},
	urldate = {2025-05-27},
	journal = {Phys. Rev. X},
	author = {Chen, Lebing and Chung, Jae-Ho and Gao, Bin and Chen, Tong and Stone, Matthew B. and Kolesnikov, Alexander I. and Huang, Qingzhen and Dai, Pengcheng},
	month = nov,
	year = {2018},
	pages = {041028},
}

@article{riberolles_chiral_2024,
	title = {Chiral and flat-band magnetic quasiparticles in ferromagnetic and metallic kagome layers},
	volume = {15},
	copyright = {2024 The Author(s)},
	issn = {2041-1723},
	url = {https://www.nature.com/articles/s41467-024-45841-8},
	doi = {10.1038/s41467-024-45841-8},
	language = {english},
	number = {1},
	urldate = {2025-05-27},
	journal = {Nat. Commun.},
	author = {Riberolles, S. X. M. and Slade, Tyler J. and Han, Tianxiong and Li, Bing and Abernathy, D. L. and Canfield, P. C. and Ueland, B. G. and Orth, P. P. and Ke, Liqin and McQueeney, R. J.},
	month = feb,
	year = {2024},
	pages = {1592},
}

@article{joannopoulos_photonic_1997,
	title = {Photonic crystals},
	volume = {102},
	issn = {0038-1098},
	url = {https://www.sciencedirect.com/science/article/pii/S0038109896007168},
	doi = {10.1016/S0038-1098(96)00716-8},
	number = {2},
	urldate = {2025-05-27},
	journal = {Solid State Commun.},
	author = {Joannopoulos, J. D. and Villeneuve, Pierre R. and Fan, Shanhui},
	month = apr,
	year = {1997},
	pages = {165--173},
}

@article{puszkarski_magnonic_2003,
	title = {Magnonic {Crystals} — the {Magnetic} {Counterpart} of {Photonic} {Crystals}},
	volume = {94},
	issn = {1662-9779},
	url = {https://www.scientific.net/SSP.94.125},
	doi = {10.4028/www.scientific.net/SSP.94.125},
	language = {english},
	urldate = {2025-05-27},
	journal = {Solid State Sci.},
	author = {Puszkarski, H. and Krawczyk, M.},
	year = {2003},
	pages = {125--134},
}

@article{chumak_current-controlled_2009,
	title = {A current-controlled, dynamic magnonic crystal},
	volume = {42},
	issn = {0022-3727},
	url = {https://dx.doi.org/10.1088/0022-3727/42/20/205005},
	doi = {10.1088/0022-3727/42/20/205005},
	language = {english},
	number = {20},
	urldate = {2025-05-27},
	journal = {J. Phys. D: Appl. Phys.},
	author = {Chumak, A V and Neumann, T and Serga, A A and Hillebrands, B and Kostylev, M P},
	month = sep,
	year = {2009},
	pages = {205005},
}

@article{wang_nanostructured_2010,
	title = {Nanostructured {Magnonic} {Crystals} with {Size}-{Tunable} {Bandgaps}},
	volume = {4},
	issn = {1936-0851},
	url = {https://doi.org/10.1021/nn901171u},
	doi = {10.1021/nn901171u},
	number = {2},
	urldate = {2025-05-27},
	journal = {ACS Nano},
	author = {Wang, Zhi Kui and Zhang, Vanessa Li and Lim, Hock Siah and Ng, Ser Choon and Kuok, Meng Hau and Jain, Shikha and Adeyeye, Adekunle Olusola},
	month = feb,
	year = {2010},
	pages = {643--648},
}

@article{tacchi_band_2011,
	title = {Band {Diagram} of {Spin} {Waves} in a {Two}-{Dimensional} {Magnonic} {Crystal}},
	volume = {107},
	url = {https://link.aps.org/doi/10.1103/PhysRevLett.107.127204},
	doi = {10.1103/PhysRevLett.107.127204},
	number = {12},
	urldate = {2025-05-27},
	journal = {Phys. Rev. Lett.},
	author = {Tacchi, S. and Montoncello, F. and Madami, M. and Gubbiotti, G. and Carlotti, G. and Giovannini, L. and Zivieri, R. and Nizzoli, F. and Jain, S. and Adeyeye, A. O. and Singh, N.},
	month = sep,
	year = {2011},
	pages = {127204},
}

@article{krauss_two-dimensional_1996,
	title = {Two-dimensional photonic-bandgap structures operating at near-infrared wavelengths},
	volume = {383},
	copyright = {1996 Springer Nature Limited},
	issn = {1476-4687},
	url = {https://www.nature.com/articles/383699a0},
	doi = {10.1038/383699a0},
	language = {english},
	number = {6602},
	urldate = {2025-05-27},
	journal = {Nature},
	author = {Krauss, Thomas F. and Rue, Richard M. De La and Brand, Stuart},
	month = oct,
	year = {1996},
	pages = {699--702},
}

@article{chumak_magnonic_2017,
	title = {Magnonic crystals for data processing},
	volume = {50},
	issn = {0022-3727},
	url = {https://dx.doi.org/10.1088/1361-6463/aa6a65},
	doi = {10.1088/1361-6463/aa6a65},
	language = {english},
	number = {24},
	urldate = {2025-05-27},
	journal = {J. Phys. D: Appl. Phys.},
	author = {Chumak, A V and Serga, A A and Hillebrands, B},
	month = may,
	year = {2017},
	pages = {244001},
}

@article{landau_theory_1935,
	title = {On the theory of the dispersion of magnetic permeability in ferromagnetic bodies},
	volume = {8},
	journal = {Phys. Zeitsch. der Sow.},
	author = {Landau, L and Lifshits, E},
	year = {1935},
	pages = {153--169},
}

@article{gilbert_phenomenological_2004,
	title = {A phenomenological theory of damping in ferromagnetic materials},
	volume = {40},
	issn = {1941-0069},
	url = {https://ieeexplore.ieee.org/document/1353448},
	doi = {10.1109/TMAG.2004.836740},
	number = {6},
	urldate = {2025-05-27},
	journal = {IEEE Trans. Magn.},
	author = {Gilbert, T.L.},
	month = nov,
	year = {2004},
	pages = {3443--3449},
}

@book{brown_micromagnetics_1963,
	address = {New York},
	title = {Micromagnetics.},
	language = {english},
	publisher = {Interscience Publishers},
	author = {Brown, William Fuller},
	year = {1963},
	note = {Open Library ID: OL5886173M},
}

@article{vansteenkiste_design_2014,
	title = {The design and verification of {MuMax3}},
	volume = {4},
	issn = {2158-3226},
	url = {https://pubs.aip.org/adv/article/4/10/107133/584191/The-design-and-verification-of-MuMax3},
	doi = {10.1063/1.4899186},
	language = {english},
	number = {10},
	urldate = {2025-05-27},
	journal = {AIP Adv.},
	author = {Vansteenkiste, Arne and Leliaert, Jonathan and Dvornik, Mykola and Helsen, Mathias and Garcia-Sanchez, Felipe and Van Waeyenberge, Bartel},
	month = oct,
	year = {2014},
	pages = {107133},
}

@article{nikitov_spin_2001,
	title = {Spin waves in periodic magnetic structures—magnonic crystals},
	volume = {236},
	issn = {0304-8853},
	url = {https://www.sciencedirect.com/science/article/pii/S030488530100470X},
	doi = {10.1016/S0304-8853(01)00470-X},
	number = {3},
	urldate = {2025-05-27},
	journal = {J. Magn. Magn. Mater.},
	author = {Nikitov, S. A. and Tailhades, Ph. and Tsai, C. S.},
	month = nov,
	year = {2001},
	pages = {320--330},
}

@article{krawczyk_magnonic_2013,
	title = {Magnonic band structures in two-dimensional bi-component magnonic crystals with in-plane magnetization},
	volume = {46},
	issn = {0022-3727},
	url = {https://dx.doi.org/10.1088/0022-3727/46/49/495003},
	doi = {10.1088/0022-3727/46/49/495003},
	language = {english},
	number = {49},
	urldate = {2025-05-27},
	journal = {J. Phys. D: Appl. Phys.},
	author = {Krawczyk, M and Mamica, S and Mruczkiewicz, M and Klos, J W and Tacchi, S and Madami, M and Gubbiotti, G and Duerr, G and Grundler, D},
	month = nov,
	year = {2013},
	pages = {495003},
}

@article{centala_compact_2023,
	title = {Compact localized states in magnonic {Lieb} lattices},
	volume = {13},
	copyright = {2023 The Author(s)},
	issn = {2045-2322},
	url = {https://www.nature.com/articles/s41598-023-39816-w},
	doi = {10.1038/s41598-023-39816-w},
	language = {english},
	number = {1},
	urldate = {2025-05-27},
	journal = {Sci. Rep.},
	author = {Centała, Grzegorz and Kłos, Jarosław W.},
	month = aug,
	year = {2023},
	pages = {12676},
}

@article{yang_flatbands_2025,
	title = {Flatbands of spin waves in two-dimensional magnonic crystals with kagome lattices},
	volume = {137},
	issn = {0021-8979},
	url = {https://doi.org/10.1063/5.0246956},
	doi = {10.1063/5.0246956},
	number = {11},
	urldate = {2025-05-27},
	journal = {J. Appl. Phys.},
	author = {Yang, Hui and Yun, Guohong and Cao, Yongjun},
	month = mar,
	year = {2025},
	pages = {113904},
}

@article{neusser_anisotropic_2010,
	title = {Anisotropic {Propagation} and {Damping} of {Spin} {Waves} in a {Nanopatterned} {Antidot} {Lattice}},
	volume = {105},
	url = {https://link.aps.org/doi/10.1103/PhysRevLett.105.067208},
	doi = {10.1103/PhysRevLett.105.067208},
	number = {6},
	urldate = {2025-05-27},
	journal = {Phys. Rev. Lett.},
	author = {Neusser, S. and Duerr, G. and Bauer, H. G. and Tacchi, S. and Madami, M. and Woltersdorf, G. and Gubbiotti, G. and Back, C. H. and Grundler, D.},
	month = aug,
	year = {2010},
	pages = {067208},
}

@article{ulrichs_magnonic_2010,
	title = {Magnonic spin-wave modes in {CoFeB} antidot lattices},
	volume = {97},
	issn = {0003-6951},
	url = {https://doi.org/10.1063/1.3483136},
	doi = {10.1063/1.3483136},
	number = {9},
	urldate = {2025-05-27},
	journal = {Appl. Phys Lett.},
	author = {Ulrichs, Henning and Lenk, Benjamin and Münzenberg, Markus},
	month = sep,
	year = {2010},
	pages = {092506},
}

@article{bali_high-symmetry_2012,
	title = {High-symmetry magnonic modes in antidot lattices magnetized perpendicular to the lattice plane},
	volume = {85},
	url = {https://link.aps.org/doi/10.1103/PhysRevB.85.104414},
	doi = {10.1103/PhysRevB.85.104414},
	number = {10},
	urldate = {2025-05-27},
	journal = {Phys. Rev. B},
	author = {Bali, R. and Kostylev, M. and Tripathy, D. and Adeyeye, A. O. and Samarin, S.},
	month = mar,
	year = {2012},
	pages = {104414},
}

@article{wang_observation_2023,
	title = {Observation of {Spin}-{Wave} {Moir\'e} {Edge} and {Cavity} {Modes} in {Twisted} {Magnetic} {Lattices}},
	volume = {13},
	url = {https://link.aps.org/doi/10.1103/PhysRevX.13.021016},
	doi = {10.1103/PhysRevX.13.021016},
	number = {2},
	urldate = {2025-05-27},
	journal = {Phys. Rev. X},
	author = {Wang, Hanchen and Madami, Marco and Chen, Jilei and Jia, Hao and Zhang, Yu and Yuan, Rundong and Wang, Yizhan and He, Wenqing and Sheng, Lutong and Zhang, Yuelin and Wang, Jinlong and Liu, Song and Shen, Ka and Yu, Guoqiang and Han, Xiufeng and Yu, Dapeng and Ansermet, Jean-Philippe and Gubbiotti, Gianluca and Yu, Haiming},
	month = apr,
	year = {2023},
	pages = {021016},
}

@incollection{stancil_magnetostatic_2009,
	address = {Boston, MA},
	title = {Magnetostatic {Modes}},
	isbn = {978-0-387-77865-5},
	url = {https://doi.org/10.1007/978-0-387-77865-5_5},
	language = {english},
	urldate = {2025-05-27},
	booktitle = {Spin {Waves}: {Theory} and {Applications}},
	publisher = {Springer US},
	author = {Stancil, Daniel D. and Prabhakar, Anil},
	year = {2009},
	doi = {10.1007/978-0-387-77865-5_5},
	pages = {139--168},
}

@incollection{stancil_quantum_2009,
	address = {Boston, MA},
	title = {Quantum {Theory} of {Spin} {Waves}},
	isbn = {978-0-387-77865-5},
	url = {https://doi.org/10.1007/978-0-387-77865-5_2},
	language = {english},
	urldate = {2025-05-27},
	booktitle = {Spin {Waves}: {Theory} and {Applications}},
	publisher = {Springer US},
	author = {Stancil, Daniel D. and Prabhakar, Anil},
	year = {2009},
	doi = {10.1007/978-0-387-77865-5_2},
	pages = {33--66},
}

@article{gros_phase_2021,
	title = {Phase resolved observation of spin wave modes in antidot lattices},
	volume = {118},
	issn = {0003-6951},
	url = {https://doi.org/10.1063/5.0045142},
	doi = {10.1063/5.0045142},
	number = {23},
	urldate = {2025-05-27},
	journal = {Appl. Phys Lett.},
	author = {Groß, Felix and Zelent, Mateusz and Gangwar, Ajay and Mamica, Sławomir and Gruszecki, Paweł and Werner, Matthias and Schütz, Gisela and Weigand, Markus and Goering, Eberhard J. and Back, Christian H. and Krawczyk, Maciej and Gräfe, Joachim},
	month = jun,
	year = {2021},
	pages = {232403},
}

@article{lim_ferromagnetic_2021,
	title = {Ferromagnetic {Resonance} {Modes} in the {Exchange}-{Dominated} {Limit} in {Cylinders} of {Finite} {Length}},
	volume = {16},
	url = {https://link.aps.org/doi/10.1103/PhysRevApplied.16.064007},
	doi = {10.1103/PhysRevApplied.16.064007},
	number = {6},
	urldate = {2025-05-27},
	journal = {Phys. Rev. Appl.},
	author = {Lim, Jinho and Garg, Anupam and Ketterson, J. B.},
	month = dec,
	year = {2021},
    pages = {064007},
}

@article{demokritov_boseeinstein_2006,
	title = {Bose–{Einstein} condensation of quasi-equilibrium magnons at room temperature under pumping},
	volume = {443},
	copyright = {2006 Springer Nature Limited},
	issn = {1476-4687},
	url = {https://www.nature.com/articles/nature05117},
	doi = {10.1038/nature05117},
	language = {english},
	number = {7110},
	urldate = {2025-05-27},
	journal = {Nature},
	author = {Demokritov, S. O. and Demidov, V. E. and Dzyapko, O. and Melkov, G. A. and Serga, A. A. and Hillebrands, B. and Slavin, A. N.},
	month = sep,
	year = {2006},
	pages = {430--433},
}

@article{tiberkevich_excitation_2019,
	title = {Excitation of coherent second sound waves in a dense magnon gas},
	volume = {9},
	copyright = {2019 The Author(s)},
	issn = {2045-2322},
	url = {https://www.nature.com/articles/s41598-019-44956-z},
	doi = {10.1038/s41598-019-44956-z},
	language = {english},
	number = {1},
	urldate = {2025-05-27},
	journal = {Sci. Rep.},
	author = {Tiberkevich, V. and Borisenko, I. V. and Nowik-Boltyk, P. and Demidov, V. E. and Rinkevich, A. B. and Demokritov, S. O. and Slavin, A. N.},
	month = jun,
	year = {2019},
	pages = {9063},
}

@article{wang_quantum_2023,
	title = {Quantum states and intertwining phases in kagome materials},
	volume = {5},
	copyright = {2023 Springer Nature Limited},
	issn = {2522-5820},
	url = {https://www.nature.com/articles/s42254-023-00635-7},
	doi = {10.1038/s42254-023-00635-7},
	language = {english},
	number = {11},
	urldate = {2025-05-27},
	journal = {Nat. Rev. Phys.},
	author = {Wang, Yaojia and Wu, Heng and McCandless, Gregory T. and Chan, Julia Y. and Ali, Mazhar N.},
	month = nov,
	year = {2023},
	pages = {635--658},
}

@article{rhim_singular_2021,
	title = {Singular flat bands},
	volume = {6},
	issn = {null},
	url = {https://doi.org/10.1080/23746149.2021.1901606},
	doi = {10.1080/23746149.2021.1901606},
	number = {1},
	urldate = {2025-05-27},
	journal = {Adv. Phs. X},
	author = {Rhim, Jun-Won and and Yang, Bohm-Jung},
	month = jan,
	year = {2021},
	pages = {1901606},
}

@article{chen_visualizing_2023,
	title = {Visualizing the localized electrons of a kagome flat band},
	volume = {5},
	url = {https://link.aps.org/doi/10.1103/PhysRevResearch.5.043269},
	doi = {10.1103/PhysRevResearch.5.043269},
	number = {4},
	urldate = {2025-05-27},
	journal = {Phys. Rev. Res.},
	author = {Chen, Caiyun and Zheng, Jiangchang and Yu, Ruopeng and Sankar, Soumya and Law, Kam Tuen and Po, Hoi Chun and Jäck, Berthold},
	month = dec,
	year = {2023},
	pages = {043269},
}

@article{multer_imaging_2023,
	title = {Imaging real-space flat band localization in kagome magnet {FeSn}},
	volume = {4},
	copyright = {2022 The Author(s)},
	issn = {2662-4443},
	url = {https://www.nature.com/articles/s43246-022-00328-1},
	doi = {10.1038/s43246-022-00328-1},
	language = {english},
	number = {1},
	urldate = {2025-05-27},
	journal = {Commun. Mater.},
	author = {Multer, Daniel and Yin, Jia-Xin and Hossain, Md Shafayat and Yang, Xian and Sales, Brian C. and Miao, Hu and Meier, William R. and Jiang, Yu-Xiao and Xie, Yaofeng and Dai, Pengcheng and Liu, Jianpeng and Deng, Hanbin and Lei, Hechang and Lian, Biao and Zahid Hasan, M.},
	month = feb,
	year = {2023},
	pages = {1--6},
}

@article{chisnell_topological_2015,
	title = {Topological {Magnon} {Bands} in a {Kagome} {Lattice} {Ferromagnet}},
	volume = {115},
	url = {https://link.aps.org/doi/10.1103/PhysRevLett.115.147201},
	doi = {10.1103/PhysRevLett.115.147201},
	number = {14},
	urldate = {2025-05-27},
	journal = {Phys. Rev. Lett.},
	author = {Chisnell, R. and Helton, J. S. and Freedman, D. E. and Singh, D. K. and Bewley, R. I. and Nocera, D. G. and Lee, Y. S.},
	month = sep,
	year = {2015},
	pages = {147201},
}

@article{holstein_field_1940,
	title = {Field {Dependence} of the {Intrinsic} {Domain} {Magnetization} of a {Ferromagnet}},
	volume = {58},
	url = {https://link.aps.org/doi/10.1103/PhysRev.58.1098},
	doi = {10.1103/PhysRev.58.1098},
	number = {12},
	urldate = {2025-05-27},
	journal = {Phys. Rev.},
	author = {Holstein, T. and Primakoff, H.},
	month = dec,
	year = {1940},
	pages = {1098--1113},
}

@article{shindou_chiral_2013,
	title = {Chiral spin-wave edge modes in dipolar magnetic thin films},
	volume = {87},
	url = {https://link.aps.org/doi/10.1103/PhysRevB.87.174402},
	doi = {10.1103/PhysRevB.87.174402},
	number = {17},
	urldate = {2025-05-27},
	journal = {Phys. Rev. B},
	author = {Shindou, Ryuichi and Ohe, Jun-ichiro and Matsumoto, Ryo and Murakami, Shuichi and Saitoh, Eiji},
	month = may,
	year = {2013},
	pages = {174402},
}

@misc{coh_python_2022,
	title = {Python {Tight} {Binding} ({PythTB})},
	url = {https://doi.org/10.5281/zenodo.12721315},
	author = {Coh, S. and Vanderbilt, D.},
	year = {2022},
    note = {Zenodo: https://doi.org/10.5281/zenodo.12721315}
}

@article{milicevic_type-iii_2019,
	title = {Type-{III} and {Tilted} {Dirac} {Cones} {Emerging} from {Flat} {Bands} in {Photonic} {Orbital} {Graphene}},
	volume = {9},
	url = {https://link.aps.org/doi/10.1103/PhysRevX.9.031010},
	doi = {10.1103/PhysRevX.9.031010},
	number = {3},
	urldate = {2025-05-27},
	journal = {Phys. Rev. X},
	author = {Milićević, M. and Montambaux, G. and Ozawa, T. and Jamadi, O. and Real, B. and Sagnes, I. and Lemaître, A. and Le Gratiet, L. and Harouri, A. and Bloch, J. and Amo, A.},
	month = jul,
	year = {2019},
	pages = {031010},
}

@article{polini_artificial_2013,
	title = {Artificial honeycomb lattices for electrons, atoms and photons},
	volume = {8},
	copyright = {2013 Springer Nature Limited},
	issn = {1748-3395},
	url = {https://www.nature.com/articles/nnano.2013.161},
	doi = {10.1038/nnano.2013.161},
	language = {english},
	number = {9},
	urldate = {2025-05-27},
	journal = {Nat. Nanotechnol.},
	author = {Polini, Marco and Guinea, Francisco and Lewenstein, Maciej and Manoharan, Hari C. and Pellegrini, Vittorio},
	month = sep,
	year = {2013},
	pages = {625--633},
}

@article{gao_visualization_2024,
	title = {Visualization of orbital twig edge states in phononic graphene lattices},
	volume = {21},
	url = {https://link.aps.org/doi/10.1103/PhysRevApplied.21.044005},
	doi = {10.1103/PhysRevApplied.21.044005},
	number = {4},
	urldate = {2025-05-27},
	journal = {Phys. Rev. Appl.},
	author = {Gao, Feng and Peng, Yu-Gui and Xiang, Xiao and Sun, Qi-Li and Zheng, Chen and Li, Bin and Zhu, Xue-Feng},
	month = apr,
	year = {2024},
	pages = {044005},
}

@article{feilhauer_unidirectional_2023,
	title = {Unidirectional spin-wave edge modes in magnonic crystal},
	volume = {11},
	issn = {2166-532X},
	url = {https://doi.org/10.1063/5.0134099},
	doi = {10.1063/5.0134099},
	number = {2},
	urldate = {2025-05-27},
	journal = {APL Mater.},
	author = {Feilhauer, J. and Zelent, M. and Zhang, Zhiwang and Christensen, J. and Mruczkiewicz, M.},
	month = feb,
	year = {2023},
	pages = {021104},
}

@article{giovannetti_substrate-induced_2007,
	title = {Substrate-induced band gap in graphene on hexagonal boron nitride: {Ab} initio density functional calculations},
	volume = {76},
	shorttitle = {Substrate-induced band gap in graphene on hexagonal boron nitride},
	url = {https://link.aps.org/doi/10.1103/PhysRevB.76.073103},
	doi = {10.1103/PhysRevB.76.073103},
	number = {7},
	urldate = {2025-05-27},
	journal = {Phys. Rev. B},
	author = {Giovannetti, Gianluca and Khomyakov, Petr A. and Brocks, Geert and Kelly, Paul J. and van den Brink, Jeroen},
	month = aug,
	year = {2007},
	pages = {073103},
}

@article{xue_scanning_2011,
	title = {Scanning tunnelling microscopy and spectroscopy of ultra-flat graphene on hexagonal boron nitride},
	volume = {10},
	copyright = {2011 Springer Nature Limited},
	issn = {1476-4660},
	url = {https://www.nature.com/articles/nmat2968},
	doi = {10.1038/nmat2968},
	language = {english},
	number = {4},
	urldate = {2025-05-27},
	journal = {Nat. Mater.},
	author = {Xue, Jiamin and Sanchez-Yamagishi, Javier and Bulmash, Danny and Jacquod, Philippe and Deshpande, Aparna and Watanabe, K. and Taniguchi, T. and Jarillo-Herrero, Pablo and LeRoy, Brian J.},
	month = apr,
	year = {2011},
	pages = {282--285},
}

@article{jackiw_solitons_1976,
	title = {Solitons with fermion number $\frac{1}{2}$},
	volume = {13},
	url = {https://link.aps.org/doi/10.1103/PhysRevD.13.3398},
	doi = {10.1103/PhysRevD.13.3398},
	number = {12},
	urldate = {2025-05-28},
	journal = {Phys. Rev. D},
	author = {Jackiw, R. and Rebbi, C.},
	month = jun,
	year = {1976},
	pages = {3398--3409},
}

@article{lu_observation_2017,
	title = {Observation of topological valley transport of sound in sonic crystals},
	volume = {13},
	copyright = {2016 Springer Nature Limited},
	issn = {1745-2481},
	url = {https://www.nature.com/articles/nphys3999},
	doi = {10.1038/nphys3999},
	language = {english},
	number = {4},
	urldate = {2025-05-28},
	journal = {Nat. Phys.},
	author = {Lu, Jiuyang and Qiu, Chunyin and Ye, Liping and Fan, Xiying and Ke, Manzhu and Zhang, Fan and Liu, Zhengyou},
	month = apr,
	year = {2017},
	pages = {369--374},
}

@article{yang_acoustic_2018,
	title = {Acoustic valley edge states in a graphene-like resonator system},
	volume = {123},
	issn = {0021-8979},
	url = {https://doi.org/10.1063/1.5009626},
	doi = {10.1063/1.5009626},
	number = {9},
	urldate = {2025-05-28},
	journal = {J. Appl. Phys.},
	author = {Yang, Yahui and Yang, Zhaoju and Zhang, Baile},
	month = jan,
	year = {2018},
	pages = {091713},
}

@article{wang_extended_2022,
	title = {Extended topological valley-locked surface acoustic waves},
	volume = {13},
	copyright = {2022 The Author(s)},
	issn = {2041-1723},
	url = {https://www.nature.com/articles/s41467-022-29019-8},
	doi = {10.1038/s41467-022-29019-8},
	language = {english},
	number = {1},
	urldate = {2025-05-28},
	journal = {Nat. Commun.},
	author = {Wang, Ji-Qian and Zhang, Zi-Dong and Yu, Si-Yuan and Ge, Hao and Liu, Kang-Fu and Wu, Tao and Sun, Xiao-Chen and Liu, Le and Chen, Hua-Yang and He, Cheng and Lu, Ming-Hui and Chen, Yan-Feng},
	month = mar,
	year = {2022},
	pages = {1324},
}

@article{noh_observation_2018,
	title = {Observation of {Photonic} {Topological} {Valley} {Hall} {Edge} {States}},
	volume = {120},
	url = {https://link.aps.org/doi/10.1103/PhysRevLett.120.063902},
	doi = {10.1103/PhysRevLett.120.063902},
	number = {6},
	urldate = {2025-05-28},
	journal = {Phys. Rev. Lett.},
	author = {Noh, Jiho and Huang, Sheng and Chen, Kevin P. and Rechtsman, Mikael C.},
	month = feb,
	year = {2018},
	pages = {063902},
}

@article{zhang_achieving_2018,
	title = {Achieving acoustic topological valley-{Hall} states by modulating the subwavelength honeycomb lattice},
	volume = {8},
	copyright = {2018 The Author(s)},
	issn = {2045-2322},
	url = {https://www.nature.com/articles/s41598-018-35214-9},
	doi = {10.1038/s41598-018-35214-9},
	language = {english},
	number = {1},
	urldate = {2025-05-28},
	journal = {Sci. Rep.},
	author = {Zhang, Zhiwang and Cheng, Ying and Liu, Xiaojun},
	month = nov,
	year = {2018},
	pages = {16784},
}

@article{wang_ultracompact_2022,
	title = {Ultracompact topological photonic switch based on valley-vortex-enhanced high-efficiency phase shift},
	volume = {11},
	copyright = {2022 The Author(s)},
	issn = {2047-7538},
	url = {https://www.nature.com/articles/s41377-022-00993-4},
	doi = {10.1038/s41377-022-00993-4},
	language = {english},
	number = {1},
	urldate = {2025-05-28},
	journal = {Light Sci. Appl.},
	author = {Wang, Hongwei and Tang, Guojing and He, Yu and Wang, Zhen and Li, Xingfeng and Sun, Lu and Zhang, Yong and Yuan, Luqi and Dong, Jianwen and Su, Yikai},
	month = oct,
	year = {2022},
	pages = {292},
}

@article{yao_edge_2009,
	title = {Edge {States} in {Graphene}: {From} {Gapped} {Flat}-{Band} to {Gapless} {Chiral} {Modes}},
	volume = {102},
	shorttitle = {Edge {States} in {Graphene}},
	url = {https://link.aps.org/doi/10.1103/PhysRevLett.102.096801},
	doi = {10.1103/PhysRevLett.102.096801},
	number = {9},
	urldate = {2025-05-28},
	journal = {Phys. Rev. Lett.},
	author = {Yao, Wang and Yang, Shengyuan A. and Niu, Qian},
	month = mar,
	year = {2009},
	pages = {096801},
}

@article{zhang_valley_2013,
	title = {Valley {Chern} numbers and boundary modes in gapped bilayer graphene},
	volume = {110},
	url = {https://www.pnas.org/doi/full/10.1073/pnas.1308853110},
	doi = {10.1073/pnas.1308853110},
	number = {26},
	urldate = {2025-05-28},
	journal = {Proc. Natl. Acad. Sci. U.S.A.},
	author = {Zhang, Fan and MacDonald, Allan H. and Mele, Eugene J.},
	month = jun,
	year = {2013},
	pages = {10546--10551},
}

@article{hidalgo-sacoto_magnon_2020,
	title = {Magnon valley {Hall} effect in $\mathrm{CrI}_3$-based van der {Waals} heterostructures},
	volume = {101},
	url = {https://link.aps.org/doi/10.1103/PhysRevB.101.205425},
	doi = {10.1103/PhysRevB.101.205425},
	number = {20},
	urldate = {2025-05-28},
	journal = {Phys. Rev. B},
	author = {Hidalgo-Sacoto, R. and Gonzalez, R. I. and Vogel, E. E. and Allende, S. and Mella, José D. and Cardenas, C. and Troncoso, Roberto E. and Munoz, F.},
	month = may,
	year = {2020},
	pages = {205425},
}

@article{funayama_quantum_2024,
	title = {Quantum valley {Hall} effect-based topological boundaries for frequency-dependent and -independent mode energy profiles},
	volume = {7},
	copyright = {2024 The Author(s)},
	issn = {2399-3650},
	url = {https://www.nature.com/articles/s42005-024-01899-w},
	doi = {10.1038/s42005-024-01899-w},
	language = {english},
	number = {1},
	urldate = {2025-05-28},
	journal = {Commun. Phys.},
	author = {Funayama, Keita and Yatsugi, Kenichi and Iizuka, Hideo},
	month = dec,
	year = {2024},
	pages = {1--7},
}

@article{schaibley_valleytronics_2016,
	title = {Valleytronics in {2D} materials},
	volume = {1},
	copyright = {2016 Macmillan Publishers Limited},
	issn = {2058-8437},
	url = {https://www.nature.com/articles/natrevmats201655},
	doi = {10.1038/natrevmats.2016.55},
	language = {english},
	number = {11},
	urldate = {2025-05-28},
	journal = {Nat. Rev. Mater.},
	author = {Schaibley, John R. and Yu, Hongyi and Clark, Genevieve and Rivera, Pasqual and Ross, Jason S. and Seyler, Kyle L. and Yao, Wang and Xu, Xiaodong},
	month = aug,
	year = {2016},
	pages = {1--15},
}

@article{liang_dirac_2024,
	title = {Dirac points and flat bands in two-dimensional magnonic crystals with honeycomb–kagome structure},
	volume = {14},
	issn = {2158-3226},
	url = {https://doi.org/10.1063/5.0182182},
	doi = {10.1063/5.0182182},
	number = {3},
	urldate = {2025-05-28},
	journal = {AIP Adv.},
	author = {Liang, Yu and Yun, Guohong and Yang, Hui and Bai, Narsu and Cao, Yongjun},
	month = mar,
	year = {2024},
	pages = {035242},
}

@incollection{landau_magnetism_nodate,
	series = {Course of {Theoretical} {Physics}},
	title = {Magnetism},
	volume = {9},
    year = {1958},
	booktitle = {Statistical {Physics} {Part} 2},
	publisher = {Pergamon Press},
	author = {Landau, L and Lifshits, E and Pitaevskii, L},
	pages = {287},
}

@article{herring_theory_1951,
	title = {On the {Theory} of {Spin} {Waves} in {Ferromagnetic} {Media}},
	volume = {81},
	url = {https://link.aps.org/doi/10.1103/PhysRev.81.869},
	doi = {10.1103/PhysRev.81.869},
	number = {5},
	urldate = {2025-05-28},
	journal = {Phys. Rev.},
	author = {Herring, Conyers and Kittel, Charles},
	month = mar,
	year = {1951},
	pages = {869--880},
}

@article{kang_pseudo-magnetic_2021,
    title = {Pseudo-magnetic field-induced slow carrier dynamics in periodically strained graphene},
    volume = {12},
    copyright = {2021 The Author(s)},
    issn = {2041-1723},
    url = {https://www.nature.com/articles/s41467-021-25304-0},
    doi = {10.1038/s41467-021-25304-0},
    language = {english},
    number = {1},
    urldate = {2025-06-04},
    journal = {Nat. Commun.},
    author = {Kang, Dong-Ho and Sun, Hao and Luo, Manlin and Lu, Kunze and Chen, Melvina and Kim, Youngmin and Jung, Yongduck and Gao, Xuejiao and Parluhutan, Samuel Jior and Ge, Junyu and Koh, See Wee and Giovanni, David and Sum, Tze Chien and Wang, Qi Jie and Li, Hong and Nam, Donguk},
    month = aug,
    year = {2021},
    pages = {5087},
}

@article{pereira_strain_2009,
    title = {Strain {Engineering} of {Graphene}'s {Electronic} {Structure}},
    volume = {103},
    url = {https://link.aps.org/doi/10.1103/PhysRevLett.103.046801},
    doi = {10.1103/PhysRevLett.103.046801},
    number = {4},
    urldate = {2025-06-04},
    journal = {Phys. Rev. Lett.},
    author = {Pereira, Vitor M. and Castro Neto, A. H.},
    month = jul,
    year = {2009},
    pages = {046801},
}

@article{jamadi_direct_2020,
    title = {Direct observation of photonic {Landau} levels and helical edge states in strained honeycomb lattices},
    volume = {9},
    copyright = {2020 The Author(s)},
    issn = {2047-7538},
    url = {https://www.nature.com/articles/s41377-020-00377-6},
    doi = {10.1038/s41377-020-00377-6},
    language = {english},
    number = {1},
    urldate = {2025-06-04},
    journal = {Light Sci. Appl.},
    author = {Jamadi, Omar and Rozas, Elena and Salerno, Grazia and Milićević, Marijana and Ozawa, Tomoki and Sagnes, Isabelle and Lemaître, Aristide and Le Gratiet, Luc and Harouri, Abdelmounaim and Carusotto, Iacopo and Bloch, Jacqueline and Amo, Alberto},
    month = aug,
    year = {2020},
    pages = {144},
}

@article{wei_strain-engineered_2024,
    title = {Strain-engineered magnon states in two-dimensional ferromagnetic monolayers},
    volume = {6},
    url = {https://link.aps.org/doi/10.1103/PhysRevResearch.6.013210},
    doi = {10.1103/PhysRevResearch.6.013210},
    number = {1},
    urldate = {2025-06-04},
    journal = {Phys. Rev. Res.},
    author = {Wei, Bin and Zhu, Jia-Ji and Song, Yun and Chang, Kai},
    month = feb,
    year = {2024},
    pages = {013210},
}

@article{wu_flat_2007,
    title = {Flat {Bands} and {Wigner} {Crystallization} in the {Honeycomb} {Optical} {Lattice}},
    volume = {99},
    url = {https://link.aps.org/doi/10.1103/PhysRevLett.99.070401},
    doi = {10.1103/PhysRevLett.99.070401},
    number = {7},
    urldate = {2025-06-12},
    journal = {Phys. Rev. Lett.},
    author = {Wu, Congjun and Bergman, Doron and Balents, Leon and Das Sarma, S.},
    month = aug,
    year = {2007},
    pages = {070401},
}

@article{sun_magnon_2021,
    title = {Magnon {Landau} levels in the strained antiferromagnetic honeycomb nanoribbons},
    volume = {3},
    url = {https://link.aps.org/doi/10.1103/PhysRevResearch.3.043223},
    doi = {10.1103/PhysRevResearch.3.043223},
    number = {4},
    urldate = {2025-06-16},
    journal = {Phys. Rev. Res.},
    author = {Sun, Junsong and Guo, Huaiming and Feng, Shiping},
    month = dec,
    year = {2021},
    pages = {043223},
}

@article{iacocca_reconfigurable_2016,
    title = {Reconfigurable wave band structure of an artificial square ice},
    volume = {93},
    url = {https://link.aps.org/doi/10.1103/PhysRevB.93.134420},
    doi = {10.1103/PhysRevB.93.134420},
    number = {13},
    urldate = {2025-07-21},
    journal = {Phys. Rev. B},
    author = {Iacocca, Ezio and Gliga, Sebastian and Stamps, Robert L. and Heinonen, Olle},
    month = apr,
    year = {2016},
    
    pages = {134420},
}

@article{liu_orbital_2022,
    title = {Orbital design of flat bands in non-line-graph lattices via line-graph wave functions},
    volume = {105},
    url = {https://link.aps.org/doi/10.1103/PhysRevB.105.085128},
    doi = {10.1103/PhysRevB.105.085128},
    number = {8},
    urldate = {2025-07-21},
    journal = {Phys. Rev. B},
    author = {Liu, Hang and Sethi, Gurjyot and Meng, Sheng and Liu, Feng},
    month = feb,
    year = {2022},
    
    pages = {085128},
}

@misc{serha_ultra-long-living_2025,
    title = {Ultra-long-living magnons in the quantum limit},
    url = {http://arxiv.org/abs/2505.22773},
    doi = {10.48550/arXiv.2505.22773},
    urldate = {2025-07-29},
    publisher = {arXiv},
    author = {Serha, Rostyslav O. and McAllister, Kaitlin H. and Majcen, Fabian and Knauer, Sebastian and Reimann, Timmy and Dubs, Carsten and Melkov, Gennadii A. and Serga, Alexander A. and Tyberkevych, Vasyl S. and Chumak, Andrii V. and Bozhko, Dmytro A.},
    month = may,
    year = {2025},
    note = {arXiv:2505.22773 [cond-mat]},
}

@article{wang_broad-wave-vector_2024,
    title = {Broad-wave-vector spin pumping of flat-band magnons},
    volume = {21},
    url = {https://link.aps.org/doi/10.1103/PhysRevApplied.21.044024},
    doi = {10.1103/PhysRevApplied.21.044024},
    number = {4},
    urldate = {2025-08-11},
    journal = {Phys. Rev. Appl.},
    author = {Wang, Jinlong and Wang, Hanchen and Chen, Jilei and Legrand, William and Chen, Peng and Sheng, Lutong and Xia, Jihao and Lan, Guibin and Zhang, Yuelin and Yuan, Rundong and Dong, Jing and Han, Xiufeng and Ansermet, Jean-Philippe and Yu, Haiming},
    month = apr,
    year = {2024},
    
    pages = {044024},
}

@article{ding_experimental_2019,
    title = {Experimental {Demonstration} of {Acoustic} {Chern} {Insulators}},
    volume = {122},
    url = {https://link.aps.org/doi/10.1103/PhysRevLett.122.014302},
    doi = {10.1103/PhysRevLett.122.014302},
    number = {1},
    urldate = {2025-08-12},
    journal = {Phys. Rev. Lett},
    author = {Ding, Yujiang and Peng, Yugui and Zhu, Yifan and Fan, Xudong and Yang, Jing and Liang, Bin and Zhu, Xuefeng and Wan, Xiangang and Cheng, Jianchun},
    month = jan,
    year = {2019},
    
    pages = {014302},
}

@article{mangussi_multi-orbital_2020,
    title = {Multi-orbital tight binding model for cavity-polariton lattices},
    volume = {32},
    issn = {0953-8984},
    url = {https://dx.doi.org/10.1088/1361-648X/ab8524},
    doi = {10.1088/1361-648X/ab8524},
    language = {english},
    number = {31},
    urldate = {2025-08-12},
    journal = {Journal of Physics: Condensed Matter},
    author = {Mangussi, Franco and Milićević, Marijana and Sagnes, Isabelle and Gratiet, Luc Le and Harouri, Abdelmounaim and Lemaître, Aristide and Bloch, Jacqueline and Amo, Alberto and Usaj, Gonzalo},
    month = may,
    year = {2020},
    
    pages = {315402},
}

@article{qi_direct_2008,
    title = {Direct observation of the ice rule in an artificial kagome spin ice},
    volume = {77},
    url = {https://link.aps.org/doi/10.1103/PhysRevB.77.094418},
    doi = {10.1103/PhysRevB.77.094418},
    number = {9},
    urldate = {2025-08-13},
    journal = {Phys. Rev. B},
    author = {Qi, Yi and Brintlinger, T. and Cumings, John},
    month = mar,
    year = {2008},
    
    pages = {094418},
}

@article{gartside_realization_2018,
    title = {Realization of ground state in artificial kagome spin ice via topological defect-driven magnetic writing},
    volume = {13},
    copyright = {2017 The Author(s)},
    issn = {1748-3395},
    url = {https://www.nature.com/articles/s41565-017-0002-1},
    doi = {10.1038/s41565-017-0002-1},
    language = {english},
    number = {1},
    urldate = {2025-08-13},
    journal = {Nat. Nanotech. },
    author = {Gartside, Jack C. and Arroo, Daan M. and Burn, David M. and Bemmer, Victoria L. and Moskalenko, Andy and Cohen, Lesley F. and Branford, Will R.},
    month = jan,
    year = {2018},
    pages = {53--58},
}

@article{hafezi_imaging_2013,
    title = {Imaging topological edge states in silicon photonics},
    volume = {7},
    copyright = {2013 Springer Nature Limited},
    issn = {1749-4893},
    url = {https://www.nature.com/articles/nphoton.2013.274},
    doi = {10.1038/nphoton.2013.274},
    language = {english},
    number = {12},
    urldate = {2025-08-13},
    journal = {Nat. Photon.},
    author = {Hafezi, M. and Mittal, S. and Fan, J. and Migdall, A. and Taylor, J. M.},
    month = dec,
    year = {2013},
    pages = {1001--1005},
}

@article{shindou_topological_2013,
    title = {Topological chiral magnonic edge mode in a magnonic crystal},
    volume = {87},
    url = {https://link.aps.org/doi/10.1103/PhysRevB.87.174427},
    doi = {10.1103/PhysRevB.87.174427},
    number = {17},
    urldate = {2025-08-13},
    journal = {Phys. Rev. B},
    author = {Shindou, Ryuichi and Matsumoto, Ryo and Murakami, Shuichi and Ohe, Jun-ichiro},
    month = may,
    year = {2013},
    
    pages = {174427},
}

@article{maranville_characterization_2006,
    title = {Characterization of magnetic properties at edges by edge-mode dynamics},
    volume = {99},
    issn = {0021-8979},
    url = {https://doi.org/10.1063/1.2167633},
    doi = {10.1063/1.2167633},
    number = {8},
    urldate = {2025-08-26},
    journal = {J. Appl. Phys. },
    author = {Maranville, B. B. and McMichael, R. D. and Kim, S. A. and Johnson, W. L. and Ross, C. A. and Cheng, Joy Y.},
    month = apr,
    year = {2006},
    pages = {08C703},
}

@article{mcmichael_edge_2006,
    title = {Edge saturation fields and dynamic edge modes in ideal and nonideal magnetic film edges},
    volume = {74},
    url = {https://link.aps.org/doi/10.1103/PhysRevB.74.024424},
    doi = {10.1103/PhysRevB.74.024424},
    number = {2},
    urldate = {2025-08-26},
    journal = {Phys. Rev. B},
    author = {McMichael, R. D. and Maranville, B. B.},
    month = jul,
    year = {2006},
    
    pages = {024424},
}

@misc{sim_code,
    title={Code for 'Emulating 2D Materials with Magnons'},
    doi={10.13012/B2IDB-8460963_V1},
    note={Available at the Illinois Databank},
    author = {Kaman, B. and Lim, J. and Y. Liu and Hoffmann, A.}
}

@article{dubs_low_2020,
    title = {Low damping and microstructural perfection of sub-40nm-thin yttrium iron garnet films grown by liquid phase epitaxy},
    volume = {4},
    url = {https://link.aps.org/doi/10.1103/PhysRevMaterials.4.024416},
    doi = {10.1103/PhysRevMaterials.4.024416},
    number = {2},
    urldate = {2025-11-24},
    journal = {Phys. Rev. Materials},
    author = {Dubs, Carsten and Surzhenko, Oleksii and Thomas, Ronny and Osten, Julia and Schneider, Tobias and Lenz, Kilian and Grenzer, Jörg and Hübner, René and Wendler, Elke},
    month = feb,
    year = {2020},
    
    pages = {024416},
}

@article{li_epitaxial_2015,
    title = {Epitaxial patterning of nanometer-thick {$\mathrm{Y}_3\mathrm{Fe}_5\mathrm{O}_{12}$} films with low magnetic damping},
    volume = {8},
    issn = {2040-3372},
    url = {https://pubs.rsc.org/en/content/articlelanding/2016/nr/c5nr06808h},
    doi = {10.1039/C5NR06808H},
    number = {1},
    urldate = {2025-11-24},
    journal = {Nanoscale},
    author = {Li, Shaozhen and Zhang, Wei and Ding, Junjia and Pearson, John E. and Novosad, Valentine and Hoffmann, Axel},
    month = dec,
    year = {2015},
    
    pages = {388--394},
}

@article{semenoff_domain_2008,
    title = {Domain {Walls} in {Gapped} {Graphene}},
    volume = {101},
    url = {https://link.aps.org/doi/10.1103/PhysRevLett.101.087204},
    doi = {10.1103/PhysRevLett.101.087204},
    number = {8},
    urldate = {2025-11-25},
    journal = {Phys. Rev. Lett},
    author = {Semenoff, G. W. and Semenoff, V. and Zhou, Fei},
    month = aug,
    year = {2008},
    
    pages = {087204},
}

@article{jung_valley-hall_2011,
    title = {Valley-{Hall} kink and edge states in multilayer graphene},
    volume = {84},
    url = {https://link.aps.org/doi/10.1103/PhysRevB.84.075418},
    doi = {10.1103/PhysRevB.84.075418},
    number = {7},
    urldate = {2025-11-25},
    journal = {Phys. Rev. B},
    author = {Jung, Jeil and Zhang, Fan and Qiao, Zhenhua and MacDonald, Allan H.},
    month = aug,
    year = {2011},
    
    pages = {075418},
}

@article{lee_gapped_2022,
    title = {Gapped edge states and quantum valley {Hall} effect in a planar honeycomb monolayer of group {III}–{V} binary compounds of the form {BX} ({X} = {N}, {P}, and {As})},
    volume = {170},
    issn = {0022-3697},
    url = {https://www.sciencedirect.com/science/article/pii/S0022369722003699},
    doi = {10.1016/j.jpcs.2022.110946},
    urldate = {2025-11-25},
    journal = {Journal of Physics and Chemistry of Solids},
    author = {Lee, Kyu Won and Lee, Cheol Eui},
    month = nov,
    year = {2022},
    pages = {110946},
}

@article{zhu_design_2018,
    title = {Design and experimental observation of valley-{Hall} edge states in diatomic-graphene-like elastic waveguides},
    volume = {97},
    url = {https://link.aps.org/doi/10.1103/PhysRevB.97.174301},
    doi = {10.1103/PhysRevB.97.174301},
    number = {17},
    urldate = {2025-11-25},
    journal = {Phys. Rev. B},
    author = {Zhu, Hongfei and Liu, Ting-Wei and Semperlotti, Fabio},
    month = may,
    year = {2018},
    
    pages = {174301},
}

@article{lee_quantum_2020,
    title = {Quantum valley {Hall} effect in wide-gap semiconductor {SiC} monolayer},
    volume = {10},
    copyright = {2020 The Author(s)},
    issn = {2045-2322},
    url = {https://www.nature.com/articles/s41598-020-61906-2},
    doi = {10.1038/s41598-020-61906-2},

    number = {1},
    urldate = {2025-11-25},
    journal = {Scientific Reports},
    author = {Lee, Kyu Won and Lee, Cheol Eui},
    month = mar,
    year = {2020},
    pages = {5044},
}

@misc{Supplemental,
note={See Supplemental Material at [URL] for extended calculations, details, justifications, and explanations.},
}

\end{document}



\title{\textbf{Supplemental Material}}
\author{Bobby Kaman}\email{Contact Author: kaman3@illinois.edu}
\affiliation{Department of Materials Science and Engineering and Materials Research Laboratory, The Grainger College of Engineering, University of Illinois Urbana-Champaign,  Urbana, Illinois 61801, USA}
\author{Jinho Lim}
\affiliation{Department of Materials Science and Engineering and Materials Research Laboratory, The Grainger College of Engineering, University of Illinois Urbana-Champaign,  Urbana, Illinois 61801, USA}
\author{Yingkai Liu}
\affiliation{Department of Physics and Institute for Condensed Matter Theory, The Grainger College of Engineering, University of Illinois Urbana-Champaign, Urbana, Illinois 61801, USA}

\author{Axel Hoffmann}\email{Contact Author: axelh@illinois.edu}
\affiliation{Department of Materials Science and Engineering and Materials Research Laboratory, The Grainger College of Engineering, University of Illinois Urbana-Champaign,  Urbana, Illinois 61801, USA}
\date{\today}

\maketitle
\section{Ground state and Geometry}\label{sec:S_Geometry}
For all micromagnetic simulations, to simulate the properties of YIG, we choose a magnetization $M_{S}=\SI{140}{\kilo  \ampere / \meter}$ and a Gilbert damping of $\alpha=10^{-4}$. Discretization cells are approximately $11\times11\times15$ nm. These numbers are chosen carefully to represent the hexagonal lattice well with a finite number of pixels, and are numerically equal to $\left(\dfrac{100}{9},\dfrac{250}{39}\times\sqrt{3},15\right)$ nm. Time is evolved using \texttt{MuMax3}'s built-in Runge–Kutta–Fehlberg solver, with a timestep that is not fixed, but varies such that each step has an error no greater than $10^{-11}$.

It is asserted that, for 15-nm YIG thin films, in the given magnetic field and frequency range, the approximation of 1 micromagnetic cell in the $\hat{z}$ direction is a good approximation. Here we justify this. For a single magnonic crystal unit cell of \SI{15}{\nano\meter} YIG in the $d/a=0.8$ geometry under periodic boundary conditions, the resonant modes are extracted using a procedure similar to that in the main text. This is done for a $\hat{z}$-resolution of $N_z=1,10$ micromagnetic cells under $B_{\text{ext}}=180 \text{ mT }\hat{z}$. The approximation can be considered good if the spectra are the same. These are both plotted in Fig. \ref{fig:Zcell}.
\begin{figure}[h]
    \centering
    \includegraphics[width=1\linewidth]{Supp_zcell.jpg}
    \caption{Simulations of the response of single unit cells(inset) under different $\hat{z}$-resolutions.}
    \label{fig:Zcell}
\end{figure}
Perpendicular standing spin wave modes lie far above the relevant frequency range. The fact that one cell is good enough is tremendously convenient; more cells would make simulations of the desired wavevector resolution computationally intensive to perform and analyze.

\section{More band structures}\label{sec:S_MoreBandStructures}
From the plots in the main text, it is not immediately clear how the transition from small holes to large holes affects the band structure. Fig. \ref{fig:Supp_extendedBandStructures} is a series of plots meant to show this transition in the same style as the main text.

\begin{figure}
    \centering
    \includegraphics[width=1\linewidth]{Supp_0.4-0.9_Xproj.png}
    \caption{A series of band structures resembling those in the main text, but for different hole diameter to lattice parameter ratios $d/a$ ($a$ is still fixed to \SI{333}{\nano\meter}). These are projections of the response into the $(k_x,f)$ plane.}
    \label{fig:Supp_extendedBandStructures}
\end{figure}
This demonstrates the increasing band flattening as a function of $d/a$ ratio as well as the increasing accuracy of TB, especially in the higher bands near \SI{4}{\giga\hertz}.

\section{Dipole induced gap}\label{sec:S_DipoleGap}
The Dirac points are gapped slightly by the dipole-dipole interaction. The gap is visible when plotted at a greater frequency resolution in Fig. \ref{fig:Supp_Dipolegap}, and is apparently not present in the simulations with dipole-dipole interactions excluded. This is the subject of  another study in progress; this publication focuses on the ability of tight-binding to model very thin YIG films. 
\begin{figure}
    \centering
    \includegraphics[width=0.5\linewidth]{Supp_diracgap.png}
    \caption{Dispersion near the Dirac point in (a) \SI{15}{\nano\meter} thick thin film in the $d/a=0.8$ geometry which is discussed in the main text, under $B_{\text{ext}}=185 \text{ mT }\hat{z}$ (b) the same with the demagnetizing field artificially turned off, under $B_{\text{ext}}=5 \text{ mT }\hat{z}$}
    \label{fig:Supp_Dipolegap}
\end{figure}

\section{Schr\"odinger Equation}\label{sec:S_Schrodinger}
The main text claims that the interesting features of the band structure have little to do with spin waves because the Schr\"odinger equation can yield similar features in a similar geometry. This means many of the findings reported here are generalizable, and should exist, for example, in engineered electronic potentials. Here, we prepared a python script which takes a grayscale image representing a unit cell's potential energy landscape, and solves the Schr\"odinger equation $E\psi(\mathbf{r})=\mathcal{H}\psi(\mathbf{r}) =\left(V(\mathbf{r})-\dfrac{\nabla^2}{2m}\right)\psi(\mathbf{r})$ in a finite-element way for different momenta to map out a dispersion relation. This involves writing $\mathcal{H}$ as a matrix where $\mathcal{H}_{ii}$ is given by the potential, i.e. the grayscale value pulled from the image, and $\mathcal{H}_{ij}$ is calculated by the discrete version of $\nabla^2$. The matrix can then be diagonalized to yield energies. For nonzero momenta, the boundary conditions are not perfectly periodic, but instead acquire a Bloch phase $\psi(\mathbf{r+\delta\mathbf{r}})=e^{i\mathbf{k}\cdot\delta\mathbf{r}}\psi({\mathbf{r}})$. The diagonalization can then be repeated while varying the 2D vector $\mathbf{k}$ to map out a disperion relation. To confine the electrons to the "film," the chosen electronic potential is an antidot lattice with a high potential outside the antidot geometry. With unitless parameters for simplicity, the mass is set to $1$ and the potential height $|V|$ is set to $600$ -- the goal is \textit{not} to make good predictions using this method, but to demonstrate that seemingly different physics can yield similar results. To point out one issue, this implementation yields boundary conditions at edges which are different from the case of sipn waves.

\begin{figure}
    \centering
    \includegraphics[width=0.5\linewidth]{Supp_Schrodinger.png}
    \caption{Band structure from the Schrodinger equation in a $d/a=0.8$ antidot-shaped potential (inset). This is similar to the $d/a=0.8$ and $d/a=0.9$ geometries' magnonic band structures. (right) some examples of $\mathbf{\Gamma}$-point Bloch functions, which resemble the eigenmodes of magnons in the antidot lattice.}
    \label{fig:Schrodinger}
\end{figure}

The unitless dispersion for the unit cell inset is plotted in Fig \ref{fig:Schrodinger}. Differences are likely related to the dipole-dipole interaction or the boundary conditions applied in the micromagnetic simulations (which are different from those naturally enforced by a suddenly varying potential), but it is not entirely clear. For a solution to a very similar problem in the context of cold atoms, see \cite{wu_p_xy-orbital_2008}. This reference finds a similar band structure for cold atoms in a hexagonal optical lattice, where the effective atomic potential follows an anti-dot-like profile.

\section{Berry curvature calculations in inversion-broken crystals}\label{sec:S_BerryCurvature}
\subsection{Berry curvature maps}
To motivate the existence of states on the boundary between different inversion-broken phases, we examine the Berry curvature (BC) in our tight-binding models. We choose small inversion-breaking terms $\Delta\varepsilon_s$ and $\Delta\varepsilon_p$ so the curvature is concentrated near gap openings and thus has an obvious interpretation. Using \textsc{PythTB}, we calculate the BC for different signs of $\varepsilon_s$ and $\varepsilon_p$ in the relevant bands. (More precisely, we calculate the Berry flux through spaces on a dense grid; this quantity is plotted for each grid point.)

The important demonstration here is the sign of the Berry curvature at small gaps:
\begin{enumerate}
    \item Always opposite above and below each gap
    \item Opposite for different signs of $\Delta\varepsilon_s,\Delta\varepsilon_p$.
    \item Opposite for $s$ (bands 0,1) and $p$ (bands 3,4) cases
    
\end{enumerate}
This second observation motivates that a boundary between the two phases may have an edge state; the third observation motivates that edge states in the two gaps will have opposite propagation. These calculations closely follow an instructive \textsc{PythTB} example problem for graphene \cite{coh_python_2022}.
The dependence of the BC maps on the size of band gaps is not straightforward, so we also include a series of plots that show this evolution. The results are summarized in Figures \ref{fig:S_BC_Bands} and \ref{fig:BC}. Figure \ref{fig:S_BC_Bands} shows band structures as a function of symmetry-breaking parameters and \ref{fig:BC} shows the BC textures of the first 6 bands.
\begin{figure}
    \centering
    \includegraphics[width=1\linewidth]{BC_Bands.png}
    \caption{Band structures as a function of the strength of symmetry breaking. (a-b) band structures for small differently-signed parameters. (c-e) band structures for parameters increasing to approach those used in the main text. Note that no bands cross as the gap increases.}
    \label{fig:S_BC_Bands}
\end{figure}

\begin{figure}
    \centering
    \includegraphics[width=1\linewidth]{BC_Maps.png}
    \caption{Berry curvature maps. (a) band structure with labeled band indices.
    (b-c) Berry curvature for small gaps $=\pm0.2\Delta\varepsilon$ (i.e. $20\%$ of normal), demonstrating the opposite BC texture for opposite symmetry-breaking parameters.(c-e) the same for larger band gaps $(0.4, 0.6, 0.8)\Delta\varepsilon$. In the lowest bands, BC is strongly concentrated. In bands 3 and 4, the BC is strongly concentrated at the valleys for small band gaps, but it is smeared out for larger band gaps and is overshadowed by the BC texture inherited from the neighboring flat bands.}
    \label{fig:BC}
\end{figure}

\subsection{Valley Chern numbers}
Valley Chern numbers (VCNs) can be calculated by the integration of the BC in the vicinity of a valley. Because this involves a small portion of the Brillouin zone, VCNs are not true Chern numbers and are therefore not strictly quantized. However, they are still a useful diagnostic when BC is strongly localized in momentum space. For instance, the small-gap BC maps shown in Figure \ref{fig:BC} have curvature strongly localized near valleys, so the VCN can be considered a good metric of band topology. As band gaps get larger, the BC is smeared out and the VCN is a less helpful metric as the necessary region of integration becomes less well-defined. This is illustrated in Figure \ref{fig:S_VCN}, which shows the numerical VCN calculation from integration in a small region near a valley. At small gaps, the VCNs are indeed quantized to their ideal values of $\pm1/2$. At larger gaps, the VCNs lose their quantization. However, because no bands cross as the gap increases (the band structures are adiabatically connected), the bands stay in the quantum valley-Hall phase, and boundary states persist. This interpretation is supported by the existence of boundary modes in large-gap cases, as well as their change in character upon inverting the phase boundary (section \ref{sec:S_InvertedBoundary}).
\begin{figure}
    \centering
    \includegraphics[width=0.5\linewidth]{S_VCN.png}
    \caption{Valley Chern number calculations near the $\mathbf{K}$ point. For very small gaps $\Delta\varepsilon\approx0$, the VCNs are quantized to their ideal values of $\pm1/2$. The VCN picture breaks down more quickly for bands 3 and 4.}
    \label{fig:S_VCN}
\end{figure}

\section{Inverted boundary modes}\label{sec:S_InvertedBoundary}
The above section \ref{sec:S_BerryCurvature} demonstrates that the signs of Berry curvatures switch along with the sign of inversion-breaking terms $\Delta\varepsilon_s$ and $\Delta\varepsilon_p$. Naively, a boundary of inverted type should have edge states which bridge the gaps in opposite ways. This turns out to be true, and is demonstrated by a TB model in Fig. \ref{fig:Supp_invertedboundary}. 

\begin{figure}
    \centering
    \includegraphics[width=0.8\linewidth]{Supp_invertedboundary.png}
    \caption{Phase boundaries related by inversion. (a) the situation in the main text, (b) its brother, related by inversion, with an opposite sign of symmetry-breaking terms. The states at the boundary (those plotted with dark opacity) bridge the gap in opposite ways. This provides strong support for the QVH interpretation despite the breakdown of VCNs for larger gaps.}
    \label{fig:Supp_invertedboundary}
\end{figure}

\section{Backscattering of boundary modes}\label{sec:S_Backscattering}
Typically, excitations referred to as topological are seen as attractive because they are in some sense robust to imperfections - if their existence can be argued from a band topology point of view, then the states should be robust to imperfections that do not change band topology. However, if translation symmetry is broken, edge states are free scatter into each other if they share a frequency (i.e., satisfy energy conservation). It is important to note that the attractive one-way transport of Chern insulators is not present in our system. To demonstrate this, we excite Gaussian wavepackets of QVH-like states and show their scattering from a defect in Figure \ref{fig:S_Backscattering}. One honeycomb site is removed and a right-propagating state is allowed to collide with it so reflection and transmission amplitudes are calculable. Using the maximum power on either side of the defect, we define $|R|^2=\dfrac{\text{max}(|\Psi_{\text{left}} |^2)}{\text{max}(|\Psi_{\text{left}} |^2)+\text{max}(|\Psi_{\text{right}} |^2)}$. The two boundary modes experience different reflection coefficients. It is helpful to compare the real-space momentum (i.e., not the crystal momentum) of each mode to the size of the defect: the defect is large compared to $p$-mode wavelengths, but small compared to $s$-mode wavelengths.
\begin{figure}
    \centering
    \includegraphics[width=1\linewidth]{Reflection_figure.png}
    \caption{Backscattering of boundary states from a defect. (a) Spin waves are excited and move along the boundary between different gapped phases. About $\SI{20}{\micro \meter}$ away, a defect is formed by removing a single honeycomb site. $|\Psi|^2$ is plotted as a function of time and position to quantify backscattering amplitudes. (b) shows a line-cut of the power for the $s$-gap mode after the scattering event marked by the dashed line. Solid gray lines mark the position of the defect. (c) shows the same for the $p$-gap mode. Despite the defect taking up a significant portion of the boundary, the $s$-gap mode at $\SI{1.6}{\giga \hertz}$ has a small reflection, $\approx 6\%$. The $p$-gap mode at $\SI{2.7}{\giga \hertz}$ backscatters more strongly, $\approx 22 \%$. (d,e) phase-resolved plot of each snapshot, demonstrating the real-space wavelength of each mode.}
    \label{fig:S_Backscattering}
\end{figure}

\section{Defects}\label{sec:S_Defects}
To simulate the modes of point defects, the simulation field is halved in size and resolution is doubled. All real dimensions are kept the same. A single defect and a pair of defects are placed on different sides of the field to avoid unintentional coupling.
\begin{figure}
    \centering
    \includegraphics[width=0.75\linewidth]{Supp_defects.png}
    \caption{Simulation geometry. Boxed in red are the two sites of interest, with expanded versions to the right. Very small excitation regions are marked in blue. Scale bar: \SI{2}{\micro\meter}}.
    \label{fig:Supp_DefectGeom}
    
\end{figure}
\begin{figure}
    \centering
    \includegraphics[width=1\linewidth]{Supp_CoupledDefects_Response.png}
    \caption{Response $|\psi(f)|^2$ as a function of frequency of the left and the right side of the geometry of Fig. \ref{fig:Supp_DefectGeom}, showing the single (left) and coupled (right) spectra as well as the onset of bulk modes at $1.60 \text{ GHz}$. }
    \label{fig:Supp_DefectResponse}
\end{figure}
The response of the left side of the geometry clearly has a resonance at $1.47\text{ GHz}$, corresponding to the single defect-localized mode. The right side of the geometry has no such peak, but has instead one at $1.38\text{ GHz}$ and one at $1.56\text{ GHz}$. The profile $\psi(x,y,f)$ is plotted in the main text to show the nature of these two peaks. Other defect modes were resolved in this simulation, like the $p_{x,y}$-like coupled mode appearing at $1.82\text{ GHz}$, but these are of less interest because they spectrally overlap with bulk modes. 
In a tight-binding-like interpretation, localized defect states $|A\rangle$ and $|B\rangle$ are degenerate when very far away from each other, with some energy $\varepsilon$. When they are brought close together, they are coupled by some real parameter $-\Delta\equiv\langle B|\hat{H}|A\rangle$ and the new eigenmodes are given by $\dfrac{|A\rangle\pm|B\rangle}{\sqrt{2}}$ with energies $\varepsilon\mp\Delta$, resulting in a splitting of $2\Delta$. In the simulated response spectrum, the fact that the shift up- and down- in frequency is equal shows that this interpretation is good.

\section{Other TB models}\label{sec:S_OtherTB}
To show the applicability of the tight-binding type approach to other systems, we also study a square anti-dot lattice and a kagome lattice of disks. 
The kagome lattice is fairly self-explanatory, but the square lattice will be discussed a little more here.
In the same fashion as the hexagonal anti-dot lattice, we examine some Bloch-like functions. Some examples of these at the $\Gamma$ point are plotted using the same convention as in the main text.
\begin{figure}
    \centering
    \includegraphics[width=1\linewidth]{Supp_SquareBloch.png}
    \caption{Bloch-like responses of $\Gamma$-point modes on the square anti-dot lattice. These resemble a similar basis as that used in the hexagonal case (basis included in main text figure))}
    \label{fig:Supp_Squrebloch}
\end{figure}
The constants used are listed in table \ref{tab:SquareTB}. The third and fourth bands are indeed $p_{x,y}$-like, but are split into $p_x\pm ip_y$. This is due to the dipole-dipole interaction, and has been observed before, for example in \cite{lim_ferromagnetic_2021}. This phenomenon is closely related to the splitting of the Dirac point modes, and will be the subject of a following publication. The fact that the fifth mode has an apparently small response is uninteresting and is related to the exact geometry of the $\delta$-like excitation -- its overlap with this mode happens to be small.
\begin{table}
\begin{tabular}{|l|c l |}
    \hline
    $\varepsilon_s$ & 1.35 & GHz\\
    $\varepsilon _p$ & 2.34& GHz\\ 
    $\varepsilon _k$ & 2.25& GHz\\
    \hline
    $t_{sk}$ &-0.22 & GHz\\
    $t_{pk}$ & -0.29& GHz\\
    \hline
\end{tabular}
\caption{Tight-binding parameters and their values. All $\varepsilon_i$ are 'on-site' frequencies for orbital $i$ and $t_{ij}$ are hopping parameters between orbitals $i$ and $j$. Subscripts $s$ and $p$ denote $s$-like and $p_{x,y}$-like modes on the square lattice. Subscripts $k$ denote $s$-like modes between $s$ and $p$ orbitals. }
\label{tab:SquareTB}
\end{table}
It is also notable that some Bloch-like functions (one of which is plotted in Fig. \ref{fig:Supp_Squrebloch}, resemble $d$-orbitals, implying that some higher bands may also fit into a simple tight-binding like representation. For simplicity, we do not augment the model to include these. Anway, high wavevector means that these modes are likely inaccessible in experiment.

\section{Effects of inhomogeneous magnetization}\label{sec:S_Deadlayer}

One likely route to fabrication involves ion milling. Because yttrium-iron garnet is ferrimagnetic, its magnetization crucially depends on order. Ion milling locally damages the film and may lead to a magnetic dead layer. We perform simulations including an inhomogeneous saturation magnetization to mimic this scenario. The setup and results are shown in Figure \ref{fig:S_Deadlayer}. The band structure essentially resembles that of the other films, with a modified effective hole size.

\begin{figure}
    \centering
    \includegraphics[width=1\linewidth]{Deadlayer.png}
    \caption{Comparison of pristine film and film with a magnetic dead layer. (a) Spatial map of the saturation magnetization for $d/a=0.7$, as seen above in Figure \ref{fig:Supp_extendedBandStructures}. (b) Spatial map of a modified saturation magnetization meant to mimic realistic fabrication imperfections. Band structures of (c) the uniform film and (d) the film with a small magnetic dead layer $\approx30$ nm wide. This can also be compared to band structures from Figure \ref{fig:Supp_extendedBandStructures} to determine that the effects of the dead layer on the nature of spin waves is not drastic. In specific, the dead layer geometry seems to have an effective hole size between $\approx 0.6-0.7$.}
    \label{fig:S_Deadlayer}
\end{figure}

\section{Effects of disorder}\label{sec:S_Disorder}
To be experimentally relevant, the system and its attractive features must be present in the presence of realistic amounts of disorder. In a realistic fabrication scheme (for example, electron beam lithography or focused ion beam milling), the hole spacing will be consistent but edges may be distorted. To emulate this effect, we apply distortions and filters to the image that defines our 2D geometry. We then characterize these distortions in a separate analysis step. The process is as follows, and is summarized schematically in Figure \ref{fig:S_DisorderProcess}:
\begin{enumerate}
    \item For each row of pixels, shift the pixel intensity laterally by a random value in the range of $[-n,n]$ pixels.
    \item For each column of pixels, repeat the same in the vertical direction.
    \item To recover realistic edges, apply a median filter with a radius of 2 pixels.
    \item To characterize the effect this process has on the pattern, apply an edge detection filter to both the original and the distorted images. 
    \item Along line cuts through the image, measure the deviation between the position of distorted edges and original edges. The statistics of these deviations can characterize the effective linear displacement applied by this process. Dividing this deviation by the original edge spacing gives a percentage of displacement, which is a useful scalar metric of disorder.
\end{enumerate}

\begin{figure}
    \centering
    \includegraphics[width=1\linewidth]{Disorder_Process.png}
    \caption{Process for simulating random edge roughness. Measured linear displacements can be compared to average edge spacings to determine a percentage change in linear dimension. For 1px and 2px shifts, this works out to be $\approx6\%$ and $\approx11\%$ respectively.}
    \label{fig:S_DisorderProcess}
\end{figure}
\subsection{Flat bands and gaps}
The effect of disorder on inversion-broken band structures is illustrated in Figure \ref{fig:S_DisorderInvBroken}. Disorder results in a slight overall frequency shift and broadening of flat band frequencies. It's interesting to note that flat band wavefunctions are more localized than extended states, so it's possible that this could be thought of as an inhomogeneous broadening for localized wavefunctions in different environments.

\begin{figure}
    \centering
    \includegraphics[width=0.8\linewidth]{InvBroken_DisorderFigure.png}
    \caption{The effect of disorder on gapped phases. Using the process outlined above, we prepare three geometries with increasing amounts of disorder: a) $0\%$, b) $6\%$, and c) $11\%$ disorder. For each case, a representative area is shown along with the projected band structure and the total response as a function of frequency. The effect of disorder is surprisingly minimal. The band structures experience  a frequency shift of $\approx-50$ MHz and broadening particularly in the flat bands. The first flat band can be seen to increase from a FWHM of $20$ MHz at $0\%$ disorder to $80$ MHz at $11\%$ disorder. The upper flat band also experiences this broadening, reducing the effective band gap.}
    \label{fig:S_DisorderInvBroken}
\end{figure}

\subsection{Boundary modes}
The effects of disorder on boundary-localized modes are shown in Figure \ref{fig:S_Disorder_boundary}. The reciprocal space picture does not provide an intuitive demonstration of the protection or destruction of boundary-localized modes, so we also show the excitation of boundary modes for disordered geometries. The localization of these states survives disorder, but for larger amounts of disorder, there exists noticeable (inter-valley) back-scattering and excitation of other modes -- but with a magnitude that is only a small fraction of the total wavepacket power. 
\begin{figure}
    \centering
    \includegraphics[width=1\linewidth]{Boundarymodes_Disorder.png}
    \caption{Boundary modes and disorder. The same geometry studied in the main text with varying amounts of disorder: a) $0\%$, b) $6\%$, and c) $11\%$ random displacement. As expected, disorder immediately affects wavevector resolution as crystal momentum is no longer well-defined, but some states remain mostly intact. For a real-space demonstration, the $s$-like and $p$-like boundary modes are excited as a Gaussian wavepacket, and the spin wave power is shown $30$ ns after its peak. This is done for (d,e) $6\%$ disorder, with $s$-modes and $p$-modes respectively, and (f,g) $11\%$ disorder. For small disorder, small distortions to the profile are visible. This becomes more noticeable when the disorder is larger, where the trail left by the wavepacket corresponds to backscattered spin waves.}
    \label{fig:S_Disorder_boundary}

\end{figure}
\subsection{Defect modes}
The localizes modes at defects may be practically useless unless their spectral isolation is robust to disorder. To investigate this, we perform the same simulation as described in Figure \ref{fig:Supp_DefectResponse} with increasing amounts of disorder. The results are summarized in Figure \ref{fig:S_Defects_disorder}. Isolation appears to be preserved, with the effect amounting to slight frequency shifts.

\begin{figure}
    \centering
    \includegraphics[width=1\linewidth]{Defects_disorder_figure.png}
    \caption{Defect modes in the presence of disorder. The response spectrum is plotted for geometries with (a) $0$px, (b) $1$px, (c) $2$px, and (d) $5$px random shifts. Changes are minimal and seem to result only in slight frequency shifts.}
    \label{fig:S_Defects_disorder}
\end{figure}

\section{Attenuation and Quality factors}
\subsection{Attenuation lengths}
Boundary-localized spin wave transport is useless if energy dissipates over a very short distance. Here we measure the attenuation length of each boundary mode. For these simulations (along with all other simulations), we use the realistic Gilbert damping parameter of $\alpha=10^{-4}$. At an applied field $\mu_0H_z=\SI{200}{\milli \tesla}$, we excite Gaussian wavepackets of boundary-localized spin waves in the same manner as before centered around either $\SI{1.6}{\giga \hertz}$ in the $s$-band gap or $\SI{2.7}{\giga \hertz}$ in the $p$-band gap. As a metric for attenuation length, we measure the propagation distance of a wavepacket after the integrated spin wave profile has decayed to a fraction $e^{-1}$ of its maximum value. The results are shown in Figure \ref{fig:S_Attenuation}. Attenuation lengths are calculated as $\SI{63}{\micro \meter}$ and $\SI{61}{\micro \meter}$ for the $s$-gap and $p$-gap modes respectively. The fact that they are similar is a coincidence, as the two modes have different attenuation times and group velocities. These lengths are equal to about 180 unit cells, indicating that the modes are indeed expected to propagate to experimentally relevant distances.
\begin{figure}
    \centering
    \includegraphics[width=1\linewidth]{Attenuation_figure.png}
    \caption{Attenuation length of boundary-localized modes. Space-time plots of the spin wave profile are displayed next to their integrated amplitudes $|\Psi|^2$ as a function of time, for (a) the $s$-gap mode and (b) the $p$-gap mode. The simulation takes place on a ribbon with periodic boundary conditions, so each wavepacket leaves the $\pm x$ edge to return from the $\mp x$ edge. Attenuation times are given by the time taken to decay to a fraction $1/e$ of the original power, and attenuation length can be considered the propagation length during this time. These are calculated to be $\SI{63}{\micro \meter}$ and $\SI{61}{\micro \meter}$ respectively.}
    \label{fig:S_Attenuation}
\end{figure}
\subsection{Defect Q-Factors}
From an application perspective, it is also helpful to estimate quality (Q-) factors of resonances. We perform long-time simulations to resolve the frequencies of modes localized at defects to estimate quality factors. Results are shown in Figure \ref{fig:S_Defect_Qfactor}.

\begin{figure}
    \centering
    \includegraphics[width=1\linewidth]{Defect-Q.png}
    \caption{Localized resonance modes at defects. Using the same setup as before, high-resolution simulations can be used to estimate Q-factors. Amplitudes are plotted as a function of frequency near (a) the one-defect mode, (b) the acoustic coupled-defect mode, and (c) the optical coupled-defect mode.}
    \label{fig:S_Defect_Qfactor}
\end{figure}
Using the definition $Q\equiv f/\Delta f_{\text{FWHM}}$ gives $Q\approx4350,4060,$ and $ 4130$ for the one-defect, acoustic and optical modes respectively. This result is approximate due to the limited frequency resolution, and is not particularly interesting because Q- factors are essentially set by the chosen damping parameter in micromagnetic simulations. However, it is interesting to note that these linewidths are small compared to the shifts expected from disorder, for example in Figures \ref{fig:S_DisorderInvBroken} and \ref{fig:S_Defects_disorder}. Having a localized character should protect an individual mode from inhomogenous broadening.

\bibliography{references-2}